\documentclass[table]{article}
\usepackage{tikz}
\usepackage{float}
\usepackage[countmax]{subfloat}
\usetikzlibrary{trees,arrows}
\usepackage{setspace}
\usepackage{amsmath,centernot}
\usepackage{amsfonts}
\usepackage{amsthm}
\usepackage{dsfont}
\usepackage{mathtools}
\usepackage{breqn}
\usepackage{hyperref}
\usepackage{listings}
\usepackage{caption}
\usepackage[labelfont=bf]{caption}
\usepackage{array, booktabs, ragged2e}
\usepackage{makecell}
\usepackage{textpos}
\usepackage{changepage}
\usepackage{multirow}
\usepackage{graphicx}
\usepackage{bbding}
\usepackage{subcaption}
\usepackage{amssymb}
\usepackage{xcolor} 
\usepackage{circledsteps}
\usetikzlibrary{arrows.meta}
\usepackage[disable]{endfloat}  
\usepackage{threeparttable}
\usepackage{booktabs}
\usepackage{colortbl}
\usepackage{authblk}
\usepackage{easyReview}

\usepackage{csquotes}
\usepackage{titlesec}

\allowdisplaybreaks  

\definecolor{darkgray}{gray}{0.70}
\definecolor{lightgray}{gray}{0.90}

\definecolor{lblue}{rgb}{0.19, 0.55, 0.91}

\definecolor{candyapplered}{rgb}{1.0, 0.03, 0.0}

\definecolor{carnationpink}{rgb}{1.0, 0.65, 0.79}

\definecolor{limegreen}{rgb}{0.3, 0.8, 0.2}
\def\limegreen{\textcolor{limegreen}}

\definecolor{azure}{rgb}{0.0, 0.5, 1.0}
\def\azure{\textcolor{azure}}

\DeclareMathOperator{\expit}{expit}

\newcommand{\indep}{\perp \!\!\! \perp}

\usepackage[vmargin=1in,hmargin=0.9in]{geometry}

\usepackage[backend=biber,
            style=numeric-comp,
            sorting=none,
            maxnames=3,
            autocite=superscript,
            giveninits=true,
            terseinits=true]{biblatex}
\usepackage{authblk}

\title{On estimands in sequential target trial emulation}
\author[1*]{Edoardo Efrem Gervasoni}
\author[2,3]{Liesbet De Bus}
\author[1]{Stijn Vansteelandt}
\author[1]{Oliver Dukes}
\affil[1]{Department of Applied Mathematics, Computer Science and Statistics, Ghent University, Ghent, Belgium}
\affil[2]{Department of Intensive Care Medicine, Ghent University Hospital, Ghent, Belgium}
\affil[3]{Department of Internal Medicine and Pediatrics, Faculty of Medicine and Health Sciences, Ghent University, Ghent, Belgium}
\date{}

\begin{document}

\maketitle
\begin{center}
    \flushbottom{\justifying *Corresponding author email: edoardoefrem.gervasoni@ugent.be\\ Mailing address: Krijgslaan 281-S9, UGent Campus Sterre, Gent, 9000, Belgium}
\end{center}


\section*{Abstract}

The target trial framework enables causal inference from longitudinal observational data by emulating randomized trials initiated at multiple time points. Standard analyses typically pool information across trials via parametric models, defining the causal estimand implicitly as a regression coefficient.
However, this practice presents several issues: it makes the target effect dependent on the estimation strategy; it obscures interpretation when noncollapsible estimands are used; it does not reflect a clear target population when pooling across heterogeneous trials. To address these challenges, this paper proposes an estimand-first strategy for sequential target trial emulation, centered around the choice of model-free estimands, rather than model specification. This ensures that treatment effects remain clearly interpretable for well-defined populations even under model misspecification. We propose estimands suitable for different study designs, and develop accompanying G-computation and inverse probability weighted estimators. Applications on simulations and real data on antimicrobial de-escalation in an intensive care unit setting demonstrate the greater clarity and reliability of the proposed methodology over traditional techniques.

\vspace{1cm}\noindent\textbf{Keywords:} Target Trial Emulation, Causal Inference, Longitudinal Data, Semiparametric Methods


\section*{Introduction}
While ideal randomized controlled trials (RCTs) provide the most reliable causal evidence, they are often infeasible, requiring researchers to rely on observational data. However, inappropriate observational analyses can introduce avoidable biases, such as immortal-time bias.{\supercite{hernan2016immortaltime,hernan2025structural,fu2023target}}
To address these challenges, Hernán and Robins proposed the \emph{target trial framework},\supercite{hernan2016} which explicitly designs an observational study to emulate a hypothetical RCT —the \emph{target trial}—through clear definitions of its components.\supercite{fu2023target}
When eligibility criteria are met at multiple times, a sequence of target trials can be emulated, each with different baselines, and eligible patients are assigned to a treatment group according to their status at that time.
Information is then often pooled across trials using pooled Cox or logistic regression models,{\supercite{scola2023}} typically assuming, among other things, a time-constant treatment effect.
\\\\
Although appealing, these analyses raise concerns. 
Despite the existing literature on the importance of prespecifying the causal estimand before turning to estimation,{\supercite{petersen2014causal,ema2020ich,dang2023causal}} target trial analyses often rely on estimands defined implicitly as coefficients of a regression model, making the estimand dependent on the estimation strategy.{\supercite{aronow2016,stijn2022}} 
Furthermore, the well-known hazards of noncollapsibility{\supercite{apples,Hernan2010,stijn1}} acquire a new dimension in the sequential setting, as shifts in the covariate distribution across trials can produce a time-varying statistical estimand even when the underlying causal effect is constant (see later). Finally, pooling implicitly weights individuals by the number of trials they contribute to and the information each provides, potentially defeating a key objective of target trial emulation: clarity about the target population.{\supercite{keogh2023}} This issue becomes particularly acute when treatment effects are heterogeneous.
\\\\
This paper develops an estimand-first strategy for sequential target trial emulation. We propose examples of model-free estimands that do not rely on parametric modelling assumptions for their interpretation and explicitly specify how results are pooled across trials and for which target population.

\section*{Two types of sequential trials}\label{lit-rev}
Target trial emulation allows causal inference in observational studies by aligning eligibility, treatment assignment, and follow-up, thus reducing biases such as immortal time, lead time, and selection bias.\supercite{hernan2016immortaltime, fu2023target} Defining time zero is straightforward when eligibility occurs once, but becomes more complex when patients can meet eligibility criteria at multiple times. In such settings, a sequence of trials can be emulated, each with its own time zero.\supercite{hernan2008observational, gran2010sequential}
\\\\
Throughout this paper, we focus on two designs for sequential trials: \textit{visit-time} and \textit{calendar-time} trials. In visit-time trials (Fig. \textbf{1.C}), the first trial's time zero is anchored to an individual-level event, such as ICU admission, and subsequent trials are initiated at fixed intervals thereafter. This typically produces a closed cohort that progressively narrows as participants become ineligible, retaining a potentially more similar subset of patients (e.g., healthier ones) at later trials. This design is common in electronic medical records, where observation and treatment decisions are driven by patient encounters.{\supercite{covid1, al2021thrombosis, gupta2021association,massol2023levosimendan,mellado2024effect}} For example, Mathews et al.{\supercite{mathews2021prone}} emulated sequential trials among mechanically ventilated ICU patients, classifying patients as initiated on proning or not on the day of ICU admission, and repeating the process for remaining eligible patients on ICU day 2.
\begin{figure}[t]
\centering
    \includegraphics[width=1\textwidth]{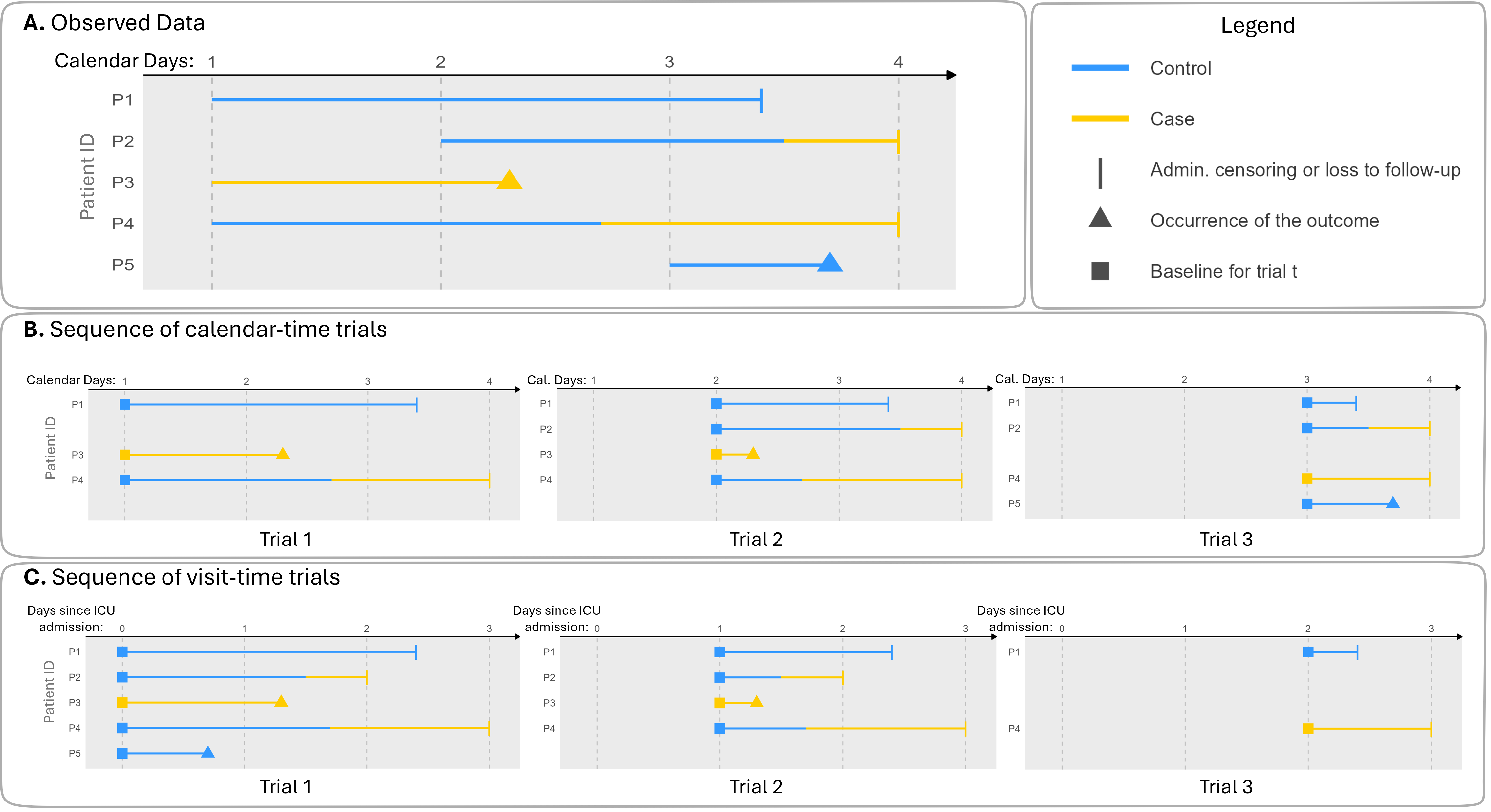}
\caption{
The figure illustrates the difference between calendar-time and visit-time sequential trials. In part \textbf{A.}, the original observed data of 5 different patients are presented, with each patient's clinical history starting at different calendar days. Part \textbf{B.} shows how the observed data can be structured as a sequence of calendar-time trials: each calendar day $t$ serves as baseline for a trial, and eligible patients at that time are classified as cases or controls. We see that newly eligible patients (P2 and P5) join trials 2 and 3. Part \textbf{C.} shows the same data organized according to a sequence of visit-time trials: now the baseline of trial 1 is given by the first occurrence of each patient in the database, e.g., the day they were admitted to the ICU. Subsequent trials have baseline shifted at $+t$ days after admission. We see that the study cohort at trial 1 progressively depletes over time, with trial 3 consisting of a subset (P1 and P4) of the eligible population of trial 1.}\vspace{0.5cm}
\end{figure}
\noindent
In contrast, calendar-time trials (Fig. \textbf{1.B}) are organized by calendar progression, reassessing eligibility at fixed intervals across the entire study population. Patients can enter and leave the study at each time point, maintaining a relatively more stable case mix. For instance, Schmidt et al.\supercite{schmidt2018diclofenac} emulated monthly trials from 1996 to 2016, rebuilding the eligible population each month. Several examples of calendar-time trials can be found in the literature.{\supercite{rossides2021, Danaei, Clark2015,caniglia2020emulating,dickerman2019avoidable,secora2020hyperkalemia,czaja2018electronic}}
While they do not cover all designs implemented in practice, visit-time and calendar-time trials were chosen to reflect common patterns in the target trial literature (Appendix S1, Table S1), and constitute the main settings for which we develop model-free causal estimands.

\section*{Notation}
\label{sec:section3}
Consider an observational study conducted over $\tau$ discrete time points, indexed by $t = 1, \dots, \tau$. At each $t$, we observe a binary treatment indicator $A_t$, a vector of time-varying covariates $L_t$, and an outcome $Y_{t}$. Outcomes may be measured with a fixed delay $\delta$ (i.e., $Y_{t+\delta}$). Let overbars denote the history of a variable, i.e., $\bar{A}_t = \{A_1, A_2, \dots, A_t\}$. We define $A_t=1$ if treatment has been initiated by time $t$, and $A_t=0$ otherwise. 
Sequential emulation involves reassessing eligibility at each time point to construct a series of trials. Individuals may contribute to multiple trials if they satisfy eligibility at those times. The specific form of the binary eligibility indicator $I_t$ may vary depending on the hypothetical trial being emulated. A common definition is to include only treatment-naïve individuals:\supercite{wang2024statistical} $I_t = 1-A_{t-1}$. Moreover, we assume that treatment has immediate effect, and define the counterfactual outcome $Y^{a}_{t}$ as the outcome that would be observed at time $t$ if, counter to the fact, an individual was assigned treatment $a$ at time $t$.
Our focus is on intention-to-treat or total effects of a point intervention.

\section*{Limitations of parametric and semiparametric estimators}
\label{subsect:parametric}

To quantify the causal effect of treatment at time $t$ on the outcome among eligible patients, one may consider the conditional contrast:
\begin{align}\label{teff}
\mathbb{E}\bigl(Y^1_{t}-Y^0_{t}\bigm|\overline{W}_t,I_t=1\bigr) \qquad \forall t,
\end{align}
where $\overline{W}_{t}$ includes the history of all possible time-dependent confounders (e.g., $\{A_1,...,A_{t-1},L_1,...,L_{t},Y_1,...,Y_{t-1}\}$). This can be readily extended to longer‐term effects of $A_t$ on subsequent outcomes. Assuming the common assumptions of consistency, positivity and sequential exchangeability, a simple strategy to estimate \eqref{teff} is to model the conditional mean of the observed outcome using a parametric linear model:
\begin{align}\label{lm}
\mathbb{E}(Y_{t}|A_t,\overline{W}_{t},I_t=1)=\beta_0+\beta_1 t+\beta^\intercal_2  \overline{W}_{t}+\psi A_t \qquad \forall t.
\end{align}
Coefficient $\psi$ represents the treatment effect. Note that this specification assumes constancy of the treatment effect across time and covariate strata. Although convenient, the OLS estimator for $\psi$ generally lacks a causal interpretation when the model is misspecified (Appendix S2.1).
The semiparametric model:
\begin{align}\label{smm}
    \mathbb{E}(Y_{t}|A_t,\overline{W}_{t},I_t=1)=\omega(t,\overline{W}_{t})+\psi A_t \qquad \forall t,
\end{align}
relaxes the assumptions of model \eqref{lm} by leaving $\omega(t,\overline{W}_{t})$ unspecified.\supercite{robinson1988root} A flexible way to estimate coefficient $\psi$ in model \eqref{smm} is via \textit{g-estimation},
which depends on correct specification of either the propensity score or the outcome model.\supercite{robins2008} Following Robins et al.,\supercite{robins1992} the \textit{g-estimator} for $\psi$ converges to:
\begin{align}
\label{g-estimator}
\frac{\mathbb{E}\biggl\{\sum^\tau_{t=1}(1-A_{t-1})Var(A_t|\overline{W}_{t},I_t=1)\mathbb{E}\left(Y^1_{t}-Y^0_{t} |\overline{W}_{t},I_t=1\right) \biggl\} }{\mathbb{E}\biggl\{\sum^\tau_{t=1}(1-A_{t-1})Var(A_t|\overline{W}_{t},I_t=1)\biggr\}},
\end{align}
where we considered $I_t=1-A_{t-1}$. 
However, g-estimation does not weight patient-times equally, as the weights $(1-A_{t-1})Var(A_t|\overline{W}_{t},I_t=1)$ in \eqref{g-estimator} favour patients who remain untreated longer or in whom treatment assignment is most uncertain. The estimated effect may thus reflect a statistically influential subgroup rather than the clinically relevant target population, complicating the interpretation of the results. This issue has been described in Keogh et al.\supercite{keogh2023} in the context of target trial emulation, and in Aronow et al.\supercite{aronow2016} more generally. This motivates the approach we implement in this paper, consisting of defining the target estimand first and only then constructing estimators for it.
Derivations of the limit expressions of the OLS and g-estimator of $\psi$ are provided in Appendices S2.1 and S2.2.

\section*{Noncollapsible effect measures}
\label{subsect:nonlinear}
Target trial analyses often rely on \textit{noncollapsible} effect measures such as the odds ratio or hazard ratio, which pose important methodological challenges.\supercite{apples, Hernan2010, stijn1} In the scoping review by Scola et al.,\supercite{scola2023} 30 of 38 studies reported the treatment effect in terms of these measures, most commonly estimated via Cox or pooled logistic regression models. 
Noncollapsibility implies that the target estimand depends on the covariates included in the model, even in the absence of confounding.\supercite{apples} As we will show next, this dependence becomes especially problematic when results are pooled across time and the distribution of treatment and covariates evolves.
\begin{figure}[t]
\centering
\scalebox{0.9}{
\begin{tikzpicture}[
    >=Stealth, 
    node distance=2cm
]
    \node (L1) at (0,10) {$L_1$};
    \node (A1) at (0,8) {$A_1$};
    \node (Y1) at (1.5,6) {$Y_1$};

    \node (L2) at (2,10) {$L_2$};
    \node (A2) at (2,8) {$A_2$};
    \node (Y2) at (3.5,6) {$Y_2$};

    \node (L3) at (4,10) {$L_3$};
    \node (A3) at (4,8) {$A_3$};
    \node (Y3) at (5.5,6) {$Y_3$};

    \node (dots1) at (6,10) {$\dots$};
    \node (dots2) at (6,8) {$\dots$};
    \node (dots3) at (7.5,6) {$\dots$};

    \draw[->] (L1) -- (Y1);
    \draw[->] (A1) -- (Y1);
    \draw[->] (A1) -- (L2);
    \draw[->] (L2) -- (Y2);
    \draw[->] (A2) -- (Y2);
    \draw[->] (A2) -- (L3);
    \draw[->] (L3) -- (Y3);
    \draw[->] (A3) -- (Y3);

\end{tikzpicture}
}
\caption{The DAG represents a data-generating process we use to illustrate the concerns of relying on noncollapsible effect measures in a time-varying setting.}
\label{fig:dag-noncollaps}\vspace{0.5cm}
\end{figure}
\\\\
\noindent
Consider a setting with randomized treatment $A_t$, binary outcome $Y_t$, and an unmeasured time-varying prognostic factor $L_t$ (e.g., frailty). 
The data-generating process is represented by the DAG in Figure \ref{fig:dag-noncollaps} and assumes a time-constant treatment effect. 
Suppose the outcome is a Bernoulli random variable with conditional mean:
\begin{align*}
    P(Y_t=1|A_t,L_t,I_t=1) = \expit(\gamma_0 + \gamma_1A_t+\gamma_2L_t) \qquad \forall t,
\end{align*}
where $\expit(h) = 1/\{1+\exp(-h)\}$. Note that the regression coefficients are time-constant. Since treatment is randomized, it is natural to estimate the effect of $A_t$ by running a marginal logistic model 
\begin{align*}
    P(Y_t=1|A_t,I_t=1) = \expit(\gamma^*_0 + \gamma^*_1A_t) \qquad \forall t
\end{align*}
over the eligible patients.
Here, $\gamma^*_0$ and $\gamma^*_1$ will generally differ from $\gamma_0$ and $\gamma_1$ since:
\begin{align*}
\expit(\gamma^*_0 + \gamma^*_1A_t) \ = \ \sum_{l_t} \expit(\gamma_0 + \gamma_1A_t+\gamma_2l_t) P(L_t=l_t|A_t,I_t=1)  \ \neq \ \expit(\gamma_0 + \gamma_1 A_t) \qquad \forall t,
\end{align*}
even in the absence of confounding.\supercite{apples} Integrals can be used in place of summations in case of continuous $L_t$. This discrepancy arises solely from the noncollapsibility of the odds ratio. Moreover, as the distribution $P(L_t=l_t|A_t,I_t=1)$ changes with time - e.g., due to selective dropout - the estimand targeted by the MLE of the marginal logistic regression becomes time-dependent, even though the underlying causal effect $\gamma_1$ is not. Because many prognostic factors are unmeasured in practice, we cannot fit a fully conditional model to mitigate this bias.
\\\\
We simulate this phenomenon in a longitudinal setting with 5 time points (details in Appendix S2.3). 
The variance of $L_t$ shrinks over time among eligible individuals, mimicking increased prognostic homogeneity. In 100 simulated datasets of size 10,000, we estimate time-specific odds ratios using marginal logistic regressions over the eligible population at each $t$.
\begin{figure}[t]
\centering
\begin{subfigure}{0.48\textwidth}
    \includegraphics[width=\textwidth]{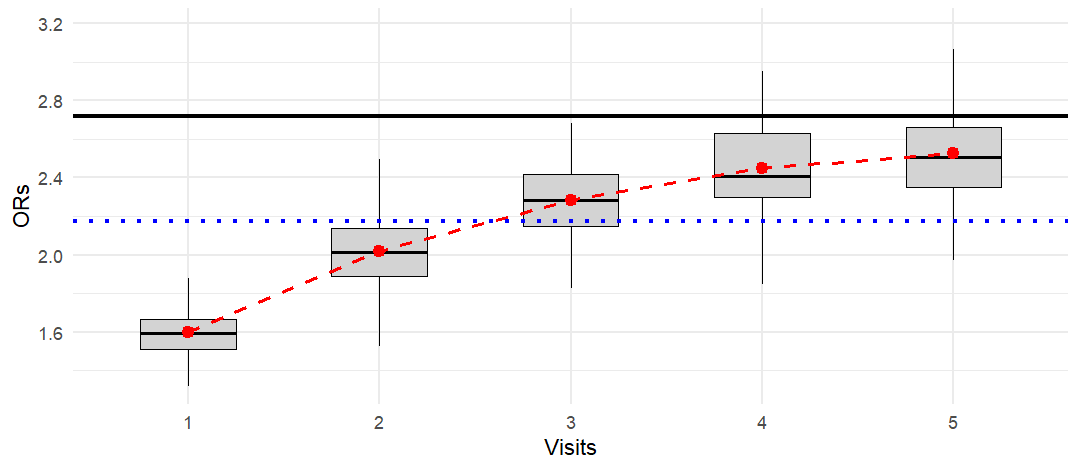}
    \caption{time-specific odds ratios}
    \label{fig:first}
\end{subfigure}
\hfill
\begin{subfigure}{0.48\textwidth}
    \includegraphics[width=\textwidth]{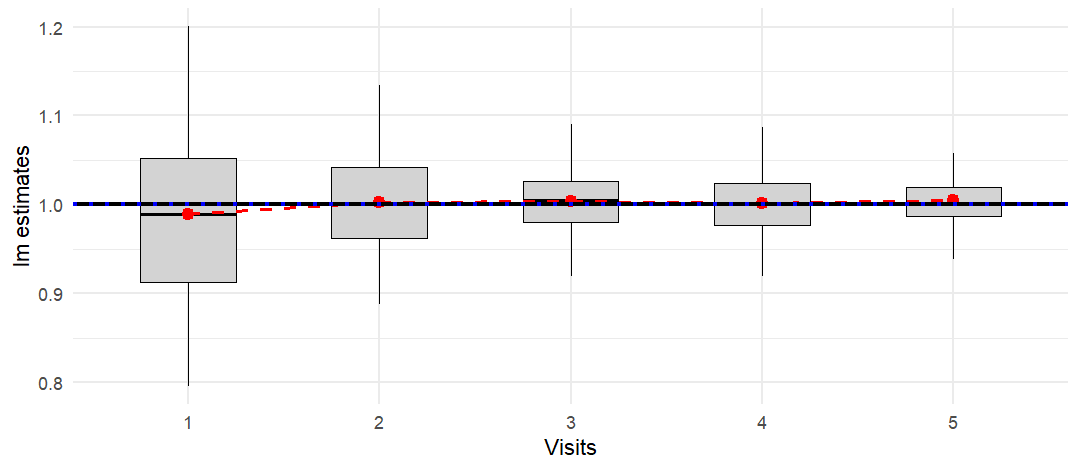}
    \caption{time-specific mean differences}
    \label{fig:second}
\end{subfigure}  
\caption{The plots show how differently noncollapsible (a) and collapsible measures (b) behave. The continuous black line represents the true conditional odds ratio $\exp(1)$ for binary $Y_t$, and the true mean difference $1$ for continuous $Y_t$. The dotted blue line represents the pooled estimate computed by respectively running a logistic regression and a linear regression over the whole dataset in long format.}\vspace{0.5cm}
\end{figure}
Figure \ref{fig:first} shows the estimated odds ratios are increasing over time, despite a constant conditional treatment coefficient of $1$ in the data-generating process. This drift is the result of the interaction between noncollapsibility and changing distribution of $L_t$, as the decreasing variance progressively brings the odds ratios closer to $\exp(1)$. This hampers interpretation of the pooled odds ratio, and obscures for what population it describes the effect.
A similar trend appears in real data (Appendix S5.3, Table S11), where time-specific odds ratios differ markedly across two visits (1.68 at $t=1$ and 5.7 at $t=2$), obscuring the causal interpretation of the estimated pooled odds ratio of 2.53 (95\% CI: 1.33–4.82). For comparison, we repeat the above experiment using a continuous outcome. In this case (Figure \ref{fig:second}), the marginal treatment effect estimates remain stable over time even with a changing distribution of $L_t$. This occurs due to the collapsibility of mean differences. 
\\\\
\noindent
An additional concern is that, unlike the OLS and g-estimators, maximum likelihood estimators of logistic regression coefficients generally lack a closed-form expression.\supercite{aronow2016} Consequently, under model misspecification, the statistical estimand targeted by the MLE lacks a transparent causal interpretation.

\section*{Model-free estimands for target trial emulation}
\label{sec:section4}
In this Section, we propose some examples of \textit{model-free estimands} for sequential trial emulation that retain a clear causal interpretation regardless of the estimation approach, and are explicit about how pooling is performed. This aligns with
the \textit{Causal Roadmap}{\supercite{dang2023causal,petersen2014causal}} and follows the principles advocated by the ICH E9 addendum.{\supercite{ema2020ich}}
\\\\
Consider sequentially emulated trials over $\tau$ time points, where each trial consists of the subset of eligible individuals at that time, denoted by $I_t = 1$. The causal contrast of interest in trial $t$ is the average treatment effect among eligible individuals:
\begin{equation*}
   \theta_t = \mathbb{E}(Y^1_t-Y_t^0|I_t=1). 
\end{equation*}
Our goal is to summarize the collection of time-specific contributions $\theta_t$ for $t=1,...,\tau$ into a single interpretable estimand in a model-free manner. We propose three such summaries, each corresponding to a different weighting scheme across time points, answering different research questions.
\begin{enumerate}
    \item \textbf{Uniformly-weighted effect in a random trial}\\
    Each trial is given equal weight, regardless of the number of eligible individuals:
    \begin{align}\label{effect_rand_trial2}
        \psi_u 
        = & \; \frac{1}{\tau}\sum_{t=1}^{\tau} \mathbb{E}(Y^1_t-Y_t^0|I_t=1).
    \end{align}
    \item \textbf{Eligibility-weighted effect in a random trial}\\
    Each trial is weighted proportionally to the fraction of eligible individuals at that time:
    \begin{align}\label{effect_rand_trial1}
        \psi_e = \sum_{t=1}^{\tau}\mathbb{E}(Y^1_t-Y_t^0|I_t=1)\frac{P(I_t=1)}{\sum_{j=1}^{\tau} P(I_j=1)}.
    \end{align}
    \item \textbf{Baseline-adjusted effect}\\
    In a visit-time design (Fig. \textbf{1.C}), the study population at $t=1$ often represents the target population, as clinicians have more control over the recruited patients, and the patient-mix is expected to be more balanced. This population gradually narrows as time progresses. It is then of interest to standardize each trial-specific contribution to the covariate distribution in that target population:
    \begin{flalign}\label{baseline-adj}
        \psi_b = \frac{1}{\tau}\sum^\tau_{t=1}\mathbb{E}\left\{\mathbb{E}\left(Y^1_{t}-Y^0_{t}|I_t=1,L_1\right)\right\}
    \end{flalign}
    so that all time-specific contrasts are interpreted relative to the fixed baseline covariate distribution $L_1$. Note that, in contrast to the calendar-time design where each trial is equally important, the progressively narrowing structure of the visit-time design gives trial 1 a central role (see Appendix S1.1). This determines a restriction on the eligibility for the estimation of $\psi_b$, that is, all patients must be eligible at baseline (details in Appendix S3.5).
\end{enumerate}

\subsection*{Interpretation of the estimands}
The proposed estimands are different summaries of trial-specific effects. They generally differ under effect modification, and the choice should be guided by which implied covariate distribution best reflects the intended target population.
\begin{figure}[ht]
\centering
\begin{subfigure}{0.4\textwidth}
    \includegraphics[width=\textwidth]{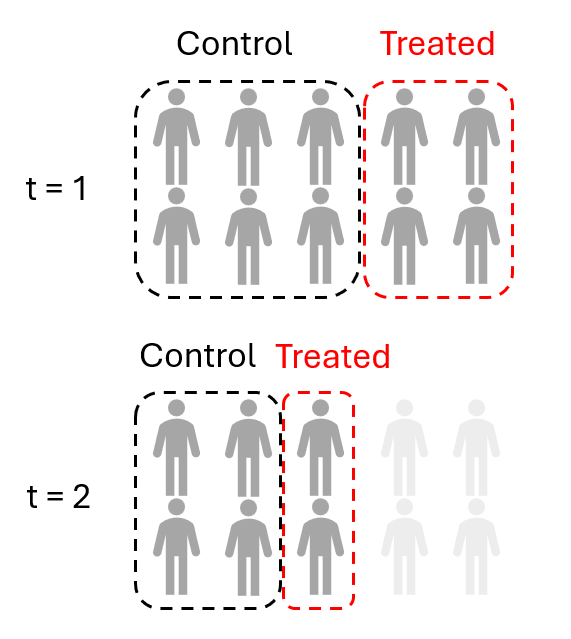}
    \caption{}
    \label{fig:ex1}
\end{subfigure}
\hfill
\centering
\begin{subfigure}{0.4\textwidth}
    \includegraphics[width=\textwidth]{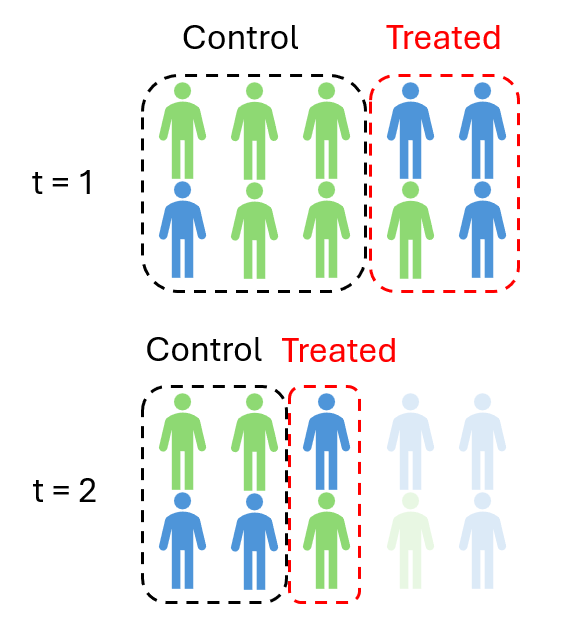}
    \caption{}
    \label{fig:ex2}
\end{subfigure}
\caption{Examples of a study with two time points. (a): 10 patients are eligible at $t=1$ and 6 at $t=2$. (b): the same example with a binary covariate $L_t$ (e.g., under mechanical ventilation in \textcolor{limegreen}{green}; not ventilated in \textcolor{azure}{azure}). At $t=1$, 6 patients are ventilated and 4 are not; at $t=2$, 3 are ventilated and 3 are not.}\vspace{0.5cm}
\end{figure}
\paragraph{Uniformly-weighted effect.}
Consider Figure~{\ref{fig:ex1}}, illustrating a study with two time points: the first trial includes 10 eligible patients and the second trial 6. Here, $\psi_u$ assigns equal weight to both trials, $\psi_u = \frac{1}{2}\theta_1 + \frac{1}{2}\theta_2$, answering the question \textit{“What is the treatment effect for the average trial (time point)?”}. This estimand suits calendar-time designs when enrollment fluctuates but no period should be overrepresented, for instance when evaluating a treatment across seasons with more respiratory admissions in winter.

\paragraph{Eligibility-weighted effect.}
From Figure~{\ref{fig:ex1}}, estimand $\psi_e = \frac{10}{16}\theta_1 + \frac{6}{16}\theta_2$ weights trials by their size, answering \textit{``What is the treatment effect for the average observation (or person-time) in the study?''} It is preferable when smaller trials contain a more selected subgroup of patients (e.g., more severe cases during ICU capacity constraints) which equal weighting would overrepresent.
\\\\
Although $\psi_u$ and $\psi_e$ were primarily developed for calendar-time data, they remain well-defined in a visit-time setting. There, $\psi_e$ is generally more interpretable, describing the effect for a randomly selected eligible patient irrespective of time since admission (details in Appendix S3.3). Moreover, the two estimands coincide when $\theta_t$ or the trial size is constant across time, though $\psi_e$ is generally estimated with greater precision due to its weighting scheme. They diverge when enrollment changes, as is common for seasonal infectious diseases or epidemic waves.
\paragraph{Baseline-adjusted effect.}
Estimand $\psi_b$ is designed for visit-time settings, retaining information from patients who drop out at later visits. In Figure~{\ref{fig:ex2}} the example is extended with a binary covariate $L_t$, mechanical ventilation: 60\% of patients are ventilated and 40\% are not at $t=1$, while at $t=2$ the two groups are evenly represented. Then
$$
\psi_b = \tfrac{1}{2}\{0.6\,\theta_1(\limegreen{\bullet}) + 0.4\,\theta_1(\azure{\bullet})\}
       + \tfrac{1}{2}\{0.6\,\theta_2(\limegreen{\bullet}) + 0.4\,\theta_2(\azure{\bullet})\},
$$
where each time-specific contribution is a linear combination of the effect within patients under mechanical ventilation ($\theta_t(\limegreen{\bullet})$) and not ($\theta_t(\azure{\bullet})$), weighted according to their distribution at baseline. This answers \textit{``What is the average effect across time for the target population identified at baseline?''} This estimand does not directly extend to calendar-time designs, as newly eligible individuals lack measurements of baseline covariates at the visit time of the initial cohort (see Appendix S3.5).

\subsection*{Identification and estimation}
\label{sec:identification}
It is possible to link the proposed estimands to observable quantities under the causal assumptions of sequential exchangeability, positivity and consistency. The identification results are in Appendix S3.5. For the uniformly-weighted effect \eqref{effect_rand_trial2}, we developed the inverse probability weighted (IPW) estimator:
\begin{align*}
\hat{\psi}_{u-ipw} = \frac{1}{\tau}\frac{1}{n}\sum_{t=1}^{\tau}\sum_{i=1}^{n}\frac{I_{t,i}}{\hat{P}(I_{t} = 1)}\biggl\{\frac{A_{t,i} Y_{t,i} }{\hat{P}(A_{t,i}=1|\overline{W}_{t,i},I_{t,i}=1)} - \frac{(1 - A_{t,i}) Y_{t,i}}{1 - \hat{P}(A_{t,i}=1|\overline{W}_{t,i},I_{t,i}=1)}\biggl\}
\end{align*}
and the G-computation estimator:
\begin{align*}
\hat{\psi}_{u-gcomp}= \frac{1}{\tau}\frac{1}{n}\sum_{t=1}^{\tau}\sum_{i=1}^{n}\frac{I_{t,i}}{\hat{P}(I_{t} = 1)}\biggl\{\hat{\mathbb{E}}(Y_{t,i}|A_{t,i} = 1,\overline{W}_{t,i},I_{t,i}=1) - \hat{\mathbb{E}}(Y_{t,i}|A_{t,i} = 0,\overline{W}_{t,i},I_{t,i}=1)\biggl\}.
\end{align*}
The corresponding IPW and G-computation estimators for the eligibility-weighted effect \eqref{effect_rand_trial1} are:
\begin{align*}
\hat{\psi}_{e-ipw} &= \frac{1}{n}\sum_{i=1}^{n}\sum_{t=1}^{\tau}\frac{I_{t,i}}{\{\sum_{j=1}^{\tau} \hat{P}(I_{j}=1)\}}\biggl\{\frac{A_{t,i} Y_{t,i} }{\hat{P}(A_{t,i}=1|\overline{W}_{t,i},I_{t,i}=1)} - \frac{(1 - A_{t,i}) Y_{t,i}}{1 - \hat{P}(A_{t,i}=1|\overline{W}_{t,i},I_{t,i}=1)}\biggl\}
\end{align*}
and 
\begin{align*}
\hat{\psi}_{e-gcomp} = \frac{1}{n}\sum_{t=1}^{\tau}\sum_{i=1}^{n}\frac{I_{t,i}}{\{\sum_{j=1}^{\tau} \hat{P}(I_{j}=1)\}}\biggl\{\hat{\mathbb{E}}(Y_{t,i}|A_{t,i} = 1,\overline{W}_{t,i},I_{t,i}=1) - \hat{\mathbb{E}}(Y_{t,i}|A_{t,i} = 0,\overline{W}_{t,i},I_{t,i}=1)\biggl\}.
\end{align*}
For the baseline-adjusted effect \eqref{baseline-adj}, we have:
\begin{equation*}
    \hat{\psi}_{b-ipw} =\frac{1}{\tau}\frac{1}{n}\sum^\tau_{t=1}\sum^n_{i=1} \frac{I_{t,i}}{\hat{P}(I_{t,i}=1|L_{1,i})}  \biggl\{\frac{A_{t,i}Y_{t,i}}{\hat{P}(A_{t,i}=1|\overline{W}_{t,i},I_{t,i}=1)} - \frac{(1-A_{t,i})Y_{t,i}}{1-\hat{P}(A_{t,i}=1|\overline{W}_{t,i},I_{t,i}=1)}\biggl\}
\end{equation*}
\newpage
and
\begin{flalign*}
    \hat{\psi}_{b-gcomp} =\frac{1}{\tau}\frac{1}{n}\sum_{t=1}^{\tau}\sum_{i=1}^{n}
    \biggl [\hat{\mathbb{E}}\left\{\hat{\mathbb{E}}(Y_{t,i}|A_{t,i}=1,I_{t,i}=1,\overline{W}_{t,i})|L_{1,i},I_{t,i}=1\right\} - \hat{\mathbb{E}}\left\{\hat{\mathbb{E}}(Y_{t,i}|A_{t,i}=0,I_{t,i}=1,\overline{W}_{t,i})|L_{1,i},I_{t,i}=1\right\}\biggl].
\end{flalign*}
The quantities marked with $\;\hat{}\;$ represent estimates of the true (conditional) probabilities and expectations. The implementation of our estimators relies on parametric working models: logistic regression for conditional probabilities and linear regression for conditional expectations (Appendix S4.2).

\section*{Simulation study}
\label{sec:simulations}
\begin{table}[b]
\centering
\begin{threeparttable}
    \caption{Summary of the simulation study.}
    \label{table:simulation}
    \renewcommand{\arraystretch}{1.2}
    \setlength{\tabcolsep}{8pt}
    \begin{tabular}{lp{10cm}}
        \rowcolor{darkgray}
        \textbf{Component} & \textbf{Description} \\
        Aims & 
        Assess performance and robustness of the proposed estimators 
        against pooled OLS under two settings: \textbf{Setting 1}, 
        time-constant treatment effect (pooled linear model correctly 
        specified); \textbf{Setting 2}, time-varying treatment effect 
        (pooled linear model misspecified). \\
        \rowcolor{lightgray}
        Data designs & 
        Calendar-time and visit-time, $\tau=2$ time points, 
        $n=1000$ sample size, $N=1000$ replications. \\
        Estimands & 
        $\psi_u$, $\psi_e$ (calendar-time); $\psi_b$ (visit-time). \\
        \rowcolor{lightgray}
        Estimators & 
        IPW and G-computation for each proposed estimand. Pooled OLS as comparators. 
        \newline
        Nuisance models: correctly specified logistic regression (propensity scores) and linear 
        regression (conditional outcome means). 
        \newline
        Standard errors via robust sandwich estimator (\texttt{geex}). \\
        Performance & 
        Empirical bias, mean SE, empirical SD\tnote{a}, 95\% CI coverage with Monte Carlo Errors. 
        \newline
        \textit{Setting} 1: bias and coverage relative to the DGP parameter 
        $\gamma_1$. 
        \newline
        \textit{Setting} 2: bias and coverage of proposed estimators 
        relative to their true values under the DGP; OLS evaluated 
        against the true values of $\psi_u$ and $\psi_e$ 
        (calendar-time) and $\psi_b$ (visit-time), as the statistical 
        estimand targeted by OLS lacks causal interpretation under 
        misspecification. \\
    \end{tabular}
     \begin{tablenotes}
         \item [a] Computed as $\sqrt{ \frac{\sum_{i=1}^N(\hat{\psi}_i-\bar{\psi})^2}{N-1}}$ with $\hat{\psi}_i$ estimate relative to the $i$-th simulation and $\bar{\psi}$ average estimate across all simulations.
     \end{tablenotes}
\end{threeparttable}
\end{table}
We designed a simulation study to assess the properties of the proposed estimators. Table {\ref{table:simulation}} summarizes the simulation study following the ADEMP structure;{\supercite{morris2019using}} Table {\ref{table:dgm}} details the two data-generating processes (DGPs) we considered: time-constant (Setting 1) and time-dependent (Setting 2) treatment effect, both with continuous $Y_t$. In Setting 1, standard parametric models such as a pooled linear regression are correctly specified and expected to perform well. In Setting 2, the linear model is misspecified. An additional scenario with binary $Y_t$ is presented in Appendix S4.1 (Tables S2--S4).

\begin{table}[b]
\centering
\begin{threeparttable}
    \caption{Summary of the data-generating process.}
    \label{table:dgm}
    \renewcommand{\arraystretch}{1.2}
    \setlength{\tabcolsep}{8pt}
    \begin{tabular}{lp{10cm}}
        \rowcolor{darkgray}
        \textbf{Component} & \textbf{Description} \\
        Variables & 
        $L_t$: time-varying confounder; $A_t$: binary treatment 
        (once treated always treated); $Y_t$: 
        continuous outcome.\\
        \rowcolor{lightgray}
        Mechanism &
        \begin{minipage}{9cm}
        \vspace{4pt}
        \begin{equation*}
        \begin{aligned}
        L_t \; &\sim \; N(\alpha_0 + \alpha_1 L_{t-1} + 
            \alpha_2 A_{t-1},\, 0.2)\\
        A_t \; &\sim \; A_{t-1} + (1-A_{t-1})\cdot \mathrm{Bernoulli}(\mathrm{expit}(
            \beta_0 + \beta_1 L_{t} + 
            \beta_2 Y_{t-1}))\\
        Y_t \; &\sim \; N(\gamma_0 + \gamma_1 A_t + \gamma_2 L_t 
            + \gamma_3 Y_{t-1},\, 0.5)
        \end{aligned}
        \end{equation*}
        \vspace{4pt}
        \end{minipage}\\
        Setting 1 & 
        $\alpha_0 = 0.5$, $\alpha_1 = 1$, $\alpha_2 = 1$; 
        \newline
        $\beta_0 = 0.5$, $\beta_1 = -1$, $\beta_2 = -1$; 
        \newline
        $\gamma_0 = 0.5$, $\gamma_1 = 1$, $\gamma_2 = 1$, 
        $\gamma_3 = 1$. \\
        \rowcolor{lightgray}
        Setting 2 & 
        All parameters as in Setting 1, except $\gamma_1 = t$. \\
    \end{tabular}
\end{threeparttable}
\end{table}
We evaluate IPW and G-computation estimators of $\psi_u$ and $\psi_e$ for calendar-time data, and corresponding estimators of $\psi_b$ for visit-time data. 
For both settings, we simulated 1000 datasets with sample size 1000 over $\tau=2$ time points. As performance measures, we report bias, average standard error (SE), empirical standard deviation (SD) and 95\% Wald confidence interval coverage, with their Monte Carlo Errors{\supercite{morris2019using}} ($\times100$). The results are compared to standard OLS estimates of the treatment effect in the pooled linear model:
\begin{align}
\label{pooled_lm}
\mathbb{E}(Y_{t,i} \mid L_{t,i}, A_{t,i}, Y_{t-1,i}) = \beta_0 + \beta_1^\top L_{t,i} + \psi A_{t,i} + \beta_2 Y_{t-1,i}
\end{align}
where $Y_{t-1,i}$ is the outcome lagged by 1 visit. The model was fitted to the simulated data in the \textit{long} (or \textit{person-time}\supercite{whatif}) format. 
Standard errors are obtained via robust \textit{sandwich} estimators using the R package \textit{geex}, to account for clones and the estimation of nuisance parameters.\supercite{geex}
\begin{table}[t]
\centering
\begin{threeparttable}
    \caption{Simulation results from Setting 1: time-fixed treatment effect and continuous outcome. Empirical Bias, Mean Standard Error, Empirical Standard Deviation, and 95\% CI Coverage of the proposed estimators and of the OLS estimator of the pooled linear model. Monte Carlo Errors are reported in parentheses and multiplied by 100.}
    \label{table1}
    \renewcommand{\arraystretch}{1.2}
    \setlength{\tabcolsep}{8pt} 
    \begin{tabular}{llcccc}
        \rowcolor{darkgray}
        Data design & Estimator & Bias & SE & SD & 95\% Coverage \\
        Calendar-time & Unif. effect IPW & 0.002 (0.22) & 0.063 (0.05) & 0.068 (0.15) & 0.947 (0.71) \\
        & Unif. effect G-comp & 0.001 (0.12) & 0.039 (0.02) & 0.038 (0.09) & 0.951 (0.68) \\
        & Elig. effect IPW & 0.0001 (0.15) & 0.045 (0.06) & 0.046 (0.10) & 0.957 (0.64) \\
        & Elig. effect G-comp & -0.002 (0.11) & 0.036 (0.01) & 0.036 (0.08) & 0.943 (0.73) \\
        & Pooled OLS & -0.001 (0.09) & 0.031 (0.002) & 0.029 (0.06) & 0.959 (0.63) \\
        \rowcolor{lightgray}
        Visit-time & Baseline-adj. IPW & -0.004 (0.22) & 0.070 (0.08) & 0.070 (0.16) & 0.946 (0.71) \\
        \rowcolor{lightgray}
        & Baseline-adj. G-comp & -0.001 (0.11) & 0.034 (0.01) & 0.035 (0.08) & 0.942 (0.74) \\
        \rowcolor{lightgray}
        & Pooled OLS & -0.001 (0.09) & 0.028 (0.002) & 0.028 (0.06) & 0.942 (0.74) \\
    \end{tabular}
    \end{threeparttable}
\end{table}
\\\\
\noindent
For Setting 1, results are presented in Table \ref{table1}. Bias and coverage are relative to the treatment effect of the data-generating process. For Setting 2, model \eqref{pooled_lm} is misspecified as it assumes a constant treatment effect, an assumption which is not guaranteed to hold in general. Results are presented in Table \ref{table2}. For each estimator, bias and coverage are reported relative to the true value of the estimand in the data-generating process, computed analytically. For the OLS estimators, bias and coverage are reported against the true value of the baseline-adjusted effect for visit-time data, and the true value of both $\psi_u$ and $\psi_e$ for calendar-time data. This is done because, under model misspecification, the target parameters of the OLS estimators do not have a clear causal interpretation, in contrast to the limits of the proposed estimators.
\begin{table}[t]
\centering
\begin{threeparttable}
    \caption{Simulation results from Setting 2: time-varying treatment effect and continuous outcome. Average estimate, Empirical Bias, Mean Standard Error, Empirical Standard Deviation, and 95\% CI Coverage of the proposed estimators and of the OLS estimator of the pooled linear model. Monte Carlo Errors are reported in parentheses and multiplied by 100.}
    \label{table2}
    \renewcommand{\arraystretch}{1.2} 
    \setlength{\tabcolsep}{8pt} 
    \begin{tabular}{llccccc}
        \rowcolor{darkgray}
        Data design & Estimator & Estimate & Bias & SE & SD & 95\% Coverage \\
        Calendar-time & Unif. effect IPW & 1.495 & -0.005 (0.24) & 0.071 (0.05) & 0.075 (0.17) & 0.948 (0.70) \\
        & Unif. effect G-comp & 1.501 & 0.001 (0.12) & 0.039 (0.03) & 0.039 (0.09) & 0.950 (0.69) \\
        & Elig. effect IPW & 1.409 & -0.003 (0.17) & 0.052 (0.08) & 0.053 (0.12) & 0.951 (0.68) \\
        & Elig. effect G-comp & 1.412 & 0.001 (0.12) & 0.036 (0.02) & 0.037 (0.08) & 0.933 (0.79) \\
        & Pooled OLS (1)\tnote{a} & 1.364 & -0.136 (0.09) & 0.031 (0.002) & 0.029 (0.06) & 0.003 (0.17) \\
        & Pooled OLS (2)\tnote{b} & & -0.047 (0.09) & & & 0.656 (1.5) \\
        \rowcolor{lightgray}
        Visit-time & Baseline-adj. IPW & 1.493 & -0.007 (0.26) & 0.079 (0.09) & 0.083 (0.19) & 0.944 (0.73) \\
        \rowcolor{lightgray}
        & Baseline-adj. G-comp & 1.498 & -0.002 (0.11) & 0.034 (0.01) & 0.034 (0.08) & 0.949 (0.70) \\
        \rowcolor{lightgray}
        & Pooled OLS & 1.221 & -0.279 (0.1) & 0.032 (0.002) & 0.033 (0.07) & 0.000 (0.0) \\
    \end{tabular}
     \begin{tablenotes}
      \footnotesize
       \item [a] Bias and Coverage are reported with respect to the true value of $\psi_u$.
       \item [b] Bias and Coverage are reported with respect to the true value of $\psi_e$.
     \end{tablenotes}
\end{threeparttable}
\end{table}
\\\\
\noindent
Overall, the proposed estimators achieved negligible bias and near-nominal coverage in both settings. 
The pooled OLS estimators had the smallest standard deviations, but are not guaranteed to unbiasedly estimate a causal effect when the linear model is misspecified.
For additional details about the simulation, refer to Appendix S4 (Figures S1 and S2).

\section*{Effect of antimicrobial de-escalation in intensive care}
\label{sec:dataanalysis}
The rise of multidrug-resistant pathogens poses a major global threat, particularly in intensive care, where broad-spectrum antimicrobials are widely prescribed.\supercite{mokrani2023antibiotic, arnold2011antibiotic} Antimicrobial de-escalation aims to limit unnecessary exposure to wide-spectrum antimicrobial drugs while maintaining effective treatment. However, the evidence on its safety and effectiveness is still inconsistent, with  many of the observational studies likely affected by immortal time and selection biases.\supercite{van2020impact, de2014impact} Studies primarily aim to assess whether de-escalation has a harmful effect on short-term endpoints, or if it has no effect.
\\\\
\noindent
We analysed an observational study based on data collected from ICU patients between 2013 and 2021. The dataset included 241 unique patients and 14 covariates, with one binary outcome, hospital survival, and three continuous outcomes: duration of hospital stay, duration of ICU stay, and total antimicrobial consumption (details in Appendix S5, Tables S5–S8). A sequence of two nested trials was emulated, producing a visit-time dataset: the first trial had baseline on ICU admission, and the second trial had baseline 24 hours later. At each baseline, eligible patients were categorized into the de-escalation (if de-escalation occurred within 24 hours of that baseline) or continuation arm (otherwise). Patients could contribute to multiple trials if eligibility was maintained; in such cases a separate record was created, with 241 eligible patients in visit 1 and 160 in visit 2. Each copy was followed from its respective baseline until death or discharge, with all time-varying covariates and outcomes updated accordingly.
\begin{table}[t]
\centering
\begin{threeparttable}
    \caption{Results of the analysis of ICU data for the 4 clinical outcomes. Estimate, Standard Error, lower and upper bounds of the 95\% Confidence Intervals.}
    \label{table5}
    \renewcommand{\arraystretch}{1.2} 
    \setlength{\tabcolsep}{8pt} 
    \begin{tabular}{llllll}
        \rowcolor{darkgray}
        Outcome & Estimator & Estimate & SE & CI lower & CI upper \\
        Hospital Survival & Unif. effect IPW &  0.035 & 0.063  &  -0.089 & 0.158\\
        & Unif. effect G-comp& 0.013 & 0.032 & -0.049 & 0.076\\
        & Elig. effect IPW& 0.041 & 0.063 & -0.082 & 0.164\\
        & Elig. effect G-comp& 0.026  & 0.038  &  -0.049 & 0.102\\
        & Baseline-adj. IPW&  0.024 & 0.059  &  -0.092 & 0.141\\
        & Baseline-adj. G-comp& 0.022 & 0.032  & -0.041 & 0.085\\
        & Pooled OLS & 0.119 & 0.048  & 0.025 & 0.213\\
        \rowcolor{lightgray}
        Hospital Time & Unif. effect IPW & -5.2  & 5.7  &-16.4 & 6.1\\
        \rowcolor{lightgray}
        & Unif. effect G-comp&  -1.6 &   3.5 & -8.5 & 5.2\\
        \rowcolor{lightgray}
        & Elig. effect IPW&  -5.3 &  5.9  & -16.8 & 6.3\\
        \rowcolor{lightgray}
        & Elig. effect G-comp& -3.3 &  4.2 & -11.5 & 4.9 \\
        \rowcolor{lightgray}
        & Baseline-adj. IPW& -3.9  & 5.2  &  -14.1 & 6.2\\
        \rowcolor{lightgray}
        & Baseline-adj. G-comp&  -2.7 &   3.5 &  -9.6 & 4.1\\
        \rowcolor{lightgray}
        & Pooled OLS & 3.2 &  4.3 & -5.3 & 11.7\\
        ICU Time & Unif. effect IPW & 0.7 &  1.9 & -3.2 &  4.6\\
        & Unif. effect G-comp& 0.0  & 1.4 & -2.7  & 2.7\\
        & Elig. effect IPW&  0.5  &  2.1  & -3.5  & 4.6\\
        & Elig. effect G-comp& 0.1  & 1.6 & -3.2 &  3.2\\
        & Baseline-adj. IPW& 0.8  &  1.9 & -2.9  & 4.5\\
        & Baseline-adj. G-comp& 0.1  &  1.4 & -2.7  & 2.7\\
        & Pooled OLS & 0.8 &  1.4 & -2.0  & 3.6\\
        \rowcolor{lightgray}
        Total AM & Unif. effect IPW & 0.9  &  2.0 &  -2.9 &  4.9\\
        \rowcolor{lightgray}
        & Unif. effect G-comp&  0.1 &  1.3 & -2.6 & 2.7\\
        \rowcolor{lightgray}
        & Elig. effect IPW&  0.7 & 2.0  &  -3.2 & 4.7\\
        \rowcolor{lightgray}
        & Elig. effect G-comp& 0.1 &  1.6 & -3.0 & 3.3\\
        \rowcolor{lightgray}
        & Baseline-adj. IPW& 1.1  &  1.9 &  -2.5 &  4.8\\
        \rowcolor{lightgray}
        & Baseline-adj. G-comp&  0.1 &  1.3 &  -2.5 & 2.8\\
        \rowcolor{lightgray}
        & Pooled OLS &  0.7 &  1.4 &  -1.9 & 3.4\\
    \end{tabular}
     \begin{tablenotes}
     \footnotesize
       \item Abbreviations: Hospital Time denotes duration of hospital stay; ICU Time denotes duration of ICU stay; Total AM denotes total antimicrobial consumption.
     \end{tablenotes}
\end{threeparttable}
\end{table}
\\\\
\noindent
We applied all proposed estimators to the ICU data for each outcome. The results are presented in Table \ref{table5}. 
Inverse probability weights for IPW estimators were truncated at the 95th percentile to reduce the influence of extreme weights.\supercite{cole2008constructing} The distributions of the estimated propensity scores are shown in Appendix S5.1 (Figure S3). For comparison, pooled OLS estimators have been applied to all the outcomes. For the binary hospital survival, the proposed estimators target risk differences, which are not directly comparable with the common pooled logistic regression. Standard errors are computed via \textit{geex} as in the simulation study. Overall, the estimated effects of de-escalation were positive but close in magnitude to the null for all outcomes except hospital time.
Almost all confidence intervals included zero, but for the continuous outcomes their width suggests that sizeable effects cannot be excluded. Standard errors were particularly large for the effect of de-escalation on duration of hospitalization, due to the presence of outliers, which may also explain the large discrepancy between IPW and G-computation estimates for this outcome. In general, G-computation delivered smaller standard errors than both IPW and pooled OLS estimators. Appendix S5.3 (Tables S10, S11) shows how the MLE estimator of the pooled logistic regression compares with the analogs of the proposed estimators on the odds ratio scale for hospital survival, showing little evidence of the effect of de-escalation. This was consistent with the results on the risk-difference scale.
\\\\
\noindent
The proposed estimators provide clinically meaningful estimates. In particular, $\psi_b$ is interpretable as the effect of de-escalation on the population of eligible patients at baseline (consisting of individuals of mean age $\approx$ 60 years, 73\% males, 83\% urgent admissions, 49\% with sepsis, 23\% bacteraemia, et cetera). Estimand $\psi_e$ reflects the effect
for the average person-time unit, who remains in care and could still undergo de-escalation throughout follow-up. Estimand $\psi_u$ reflects the effect that would be expected if each person-time had the same chance of being observed at each trial (in this setting, this is less clinically relevant since the second trial contains a more select group of patients). The target populations implied by the proposed estimands are provided in Appendix S5.2 (Table S9).
The pooled OLS estimates tended to be larger in magnitude (especially compared to G-computation), and reversed the direction for hospital time. In Table S11 of Appendix S5.3, the pooled logistic regression showed a similar pattern, producing highly inflated odds ratios, while the proposed estimators remained close to unity. Although partly driven by noncollapsibility, the results also suggest possible outcome model misspecification induced by pooling, highlighting the risks of such analyses. 

\section*{Discussion}
\label{sec:discussion}
Despite growing awareness, estimands defined as coefficients of (semi)parametric models are still reported in target trial analyses. While this approach is straightforward to implement, we argue that it is vulnerable to important concerns due to model misspecification, noncollapsibility, and heterogeneous populations across trials. Concerns about model misspecification are especially acute since standard analyses are restricted to simple models. This is because adding complexity (e.g., treatment–covariate interactions) can obscure the meaning of the target quantity. 
\\\\
\noindent
The main contribution of this paper is to address these issues by defining nonparametric estimands that retain a clear causal interpretation regardless of the data structure or the models used for estimation, and explicate how information is pooled across trials and for which  target population. This framework shifts the analytical focus from the selection of a model to the definition of clinically significant causal quantities, consistent with the principles of the \textit{Causal Roadmap}.
We proposed three estimands: the uniformly-weighted effect \eqref{effect_rand_trial2}, the eligibility-weighted effect \eqref{effect_rand_trial1}, and the baseline-adjusted effect \eqref{baseline-adj}, each motivated by a distinct causal question stemming from real-world examples. 
\\\\
\noindent
Several limitations are worth noting. First, although the proposed estimands are defined in a model-free way, estimation still involves parametric models for nuisance quantities. In practice, model misspecification could introduce bias. However, in contrast to current target trial analyses, our framework is crucially agnostic about the specific estimation techniques. This flexibility allows researchers to use arbitrarily complex models, possibly allowing for time-varying effects, thereby reducing the risk of misspecification while maintaining a clear interpretation of the target quantity. Furthermore, this approach allows for a natural extension to nonparametric estimation: in future work, we will develop estimation and inference for the proposed estimands using de-biased machine learning approaches.\supercite{hines2022demystifying,stijn2022,renson2025pulling} In this paper we focused on point interventions and short-term effects; extending this framework to dynamic treatment regimes and time-to-event outcomes is another promising direction for future research. 
Additionally, our framework assumes that treatment and covariates are registered according to a fixed discrete-time grid; complications arising from irregular observation schedules are not addressed.
Further complexities common in real-world data, such as missing data or measurement error, were not addressed in our simulations.
\\\\
\noindent
Our proposed estimands are not exhaustive. Alternative weighting schemes or conditioning sets could be used according to the target population, for instance in $\psi_b$, where each trial currently contributes equally (see Appendix S1 for an example of an alternative estimand under a different design).
Moreover, we focused on the average effect across time, but it could also be relevant to define the effect for the outcome measured $\delta$ units after baseline, in the eligible population at a certain time point, say $t=1$.
This definition would typically require additional transportability assumptions, such as assumption \textit{B4} from Dahabreh et al.\supercite{Dahabreh2020}, which may become less plausible as the study period lengthens. 
Finally, note that summary estimands sacrifice granularity regarding treatment effect heterogeneity across trials and covariate strata; researchers may additionally report trial-specific effects $\theta_t$ if effect heterogeneity is of interest.
\\\\
Our goal is to raise awareness about the limitations of conventional approaches to target trial emulation and to provide concrete examples of model-free estimands that overcome those limitations. We hope this work clarifies the advantages of a model-free perspective that separates the definition of causal effects from the choice of statistical models and encourages researchers to choose estimands reflecting their clinical objectives.


\section*{Acknowledgements}
EEG and OD would like to acknowledge support from BOF grant 202209/026 and FWO grant 1222522N. SV was supported by Advanced ERC grant ACME (101141305). The authors would like to thank Mattia Cerri and Johan Steen for help with the data management and analysis. Large language models were used to assist with proofreading and identifying typographical errors during the revision of this manuscript.

\section*{Code Availability}
All code used for simulations is made available on GitHub at \url{https://github.com/EdoardoGerva/TTE-estimands}.

\section*{Data Availability}
This study used deidentified data accessed under a data use agreement. In accordance with contractual terms and applicable privacy regulations, the data can only be shared on reasonable request to the corresponding author.


\printbibliography

\appendix
\newpage
\begin{refsection}
\setcounter{table}{0}\setcounter{figure}{0}\setcounter{equation}{0}
\renewcommand{\thetable}{S\arabic{table}}
\renewcommand{\thefigure}{S\arabic{figure}}
\renewcommand{\theequation}{M\arabic{equation}}
\renewcommand{\theHtable}{S\arabic{table}}
\renewcommand{\theHfigure}{S\arabic{figure}}
\renewcommand{\theHequation}{M\arabic{equation}}
\renewcommand{\theHsection}{S\arabic{section}}

\begin{center}
{\Large Supplementary Material for ``On estimands in sequential target trial emulation''}\\[1em]
\end{center}
\tableofcontents

\renewcommand{\thesection}{S\arabic{section}}

\section{Two types of sequential trials}
In Table \ref{table:studies}, the sequential trial studies from the scoping review by Scola et al.{\supercite{scola2023}} are listed and classified into visit-time and calendar-time designs.
\begin{table}[H]
\centering
\begin{threeparttable}
    \caption[Designs of the sequential trial studies in the scoping review by Scola et al.]{Designs of the sequential trial studies from the scoping review by Scola et al.{\supercite{scola2023}}}
    \label{table:studies}
    \renewcommand{\arraystretch}{1.2}
    \setlength{\tabcolsep}{8pt}
    \begin{tabular}{clc}
        \rowcolor{darkgray}
        \textbf{Index} & \textbf{Study} & \textbf{Design} \\
    1 & Garcia-Albéniz et al. \cite{colrect} & sequences of visit-time \\
            \rowcolor{lightgray}
        2 & Mathews et al. \cite{mathews2021prone} & visit-time \\
        3 & Al-Samkari et al. \cite{al2021thrombosis} & visit-time \\
                \rowcolor{lightgray}
        4 & Shaefi et al. \cite{covid1} & visit-time \\

        5 & Danaei et al. \cite{danaei2018electronic} & calendar-time \\
        \rowcolor{lightgray}
        6 & Rossides et al. \cite{rossides2021} & calendar-time \\

        7 & Caniglia et al. \cite{caniglia2020emulating} & calendar-time \\
        \rowcolor{lightgray}
        8 & Schmidt et al. \cite{schmidt2018diclofenac} & calendar-time \\

        9 & Dickerman et al. \cite{dickerman2019avoidable} & calendar-time \\
        \rowcolor{lightgray}
        10 & Secora et al. \cite{secora2020hyperkalemia} & calendar-time \\

        11 & Czaja et al. \cite{czaja2018electronic} & calendar-time \\
    \end{tabular}
\end{threeparttable}
\end{table}
\noindent
Garcia-Albéniz et al. implemented a more complex design of sequences of visit-time trials anchored to individual birthdays. For each age from 70 to 79, a separate sequence was emulated, with the first trial's baseline defined at each patient's birthday of that age and subsequent trials initiated at weekly intervals thereafter. This process was repeated for each birthday from age 70 to age 79, producing a total of 520 trials, with new patients potentially entering the study at each subsequent birthday. For this design, it is possible to employ a hybrid of the proposed estimands. A $\psi_b$-type estimand could first be computed within each sequence, standardizing to the covariate distribution at the corresponding birthday, and the sequence-specific effects could then be pooled through a weighted average similar to $\psi_e$ or $\psi_u$. Different weighting schemes would determine different results: weighting each sequence equally as in $\psi_u$ reflects the effect in a birthday cohort selected uniformly at random; weighting by the number of eligible patients in each sequence, as in $\psi_e$, favors the larger (and presumably earlier) cohorts. Alternatively, $\psi_u$ and $\psi_e$ could also be applied directly across all 520 trials, without explicitly considering the "sequences of visit-time trials" structure.

\subsection{The anchor event in the visit-time design}
In the progressively narrowing structure of the visit-time design, the \textit{anchor event} defining the baseline of trial 1 (e.g., admission to the ICU) plays a special role, in contrast with the calendar-time setting, where every baseline carries equal weight. When implementing a sequential visit-time design, researchers should thus define the target trial components considering the central role of the anchor event. This is well illustrated by the visit-time studies in Table~{\ref{table:studies}}; for instance, Mathews et al.{\supercite{mathews2021prone}} emulate the following eligibility criteria:
\vspace{0.1cm}
\noindent
\begin{displayquote}
\footnotesize
We included adult patients ($\geq$ 18 yr old) who had moderate-to-severe hypoxemia (Berlin criteria: Pao2/Fio2 ratio $\leq$ 200 mm Hg) (19) within the first 2 days of ICU admission while receiving invasive mechanical ventilation. [$\cdots$] We excluded patients who did not have at least one Pao2/Fio2 ratio less than or equal to 200 mm Hg in the first 2 days of ICU admission, irrespective of whether they were proned. We also excluded patients if they received extracorporeal membrane oxygenation on ICU day 1, experienced cardiac arrest or severe arrhythmia (sustained ventricular tachycardia or ventricular fibrillation) on ICU day 1, were proned prior to ICU admission, or were pregnant.
\end{displayquote}
The eligibility criteria are therefore emulated with respect to ICU admission. Criteria that ignore the anchor event's special role may instead prevent sequential emulation altogether. A further consequence of the peculiarity of trial 1 in the visit-time design is the restriction it imposes on eligibility for $\psi_b$, detailed in Appendix S3.5.


\section{Limitations of parametric and semiparametric estimators}
\subsection{Limit expression of the OLS estimator}
\label{sec:asympt-ols}
The treatment effect $\psi$ in model (2) of the main text can be estimated using ordinary least squares as: 
\[\hat{\psi}_{OLS}=\frac{\sum^n_{i=1}\sum_{t=1}^{\tau}(A_{t,i}-\hat{\alpha}_0-\hat{\alpha}_1t-\hat{\alpha}_2L_{t,i})Y_{t,i}I_{t,i}}{\sum^n_{i=1}\sum_{t=1}^{\tau}(A_{t,i}-\hat{\alpha}_0-\hat{\alpha}_1t-\hat{\alpha}_2L_{t,i})A_{t,i}I_{t,i}}\]
where $(\hat{\alpha}_0,\hat{\alpha}_1,\hat{\alpha}_2)$ are the solutions to 
\[0=\sum^n_{i=1}\sum_{t=1}^{\tau}\begin{pmatrix}1\\t\\L_{t,i}\end{pmatrix}(A_{t,i}-\alpha_0-\alpha_1t-\alpha_2L_{t,i})I_{t,i}.\]
Its population limit is:
\[\frac{\mathbb{E}\left[\sum_{t=1}^{\tau}\{A_{t} - \tilde{\pi}(L_t)\}Y_{t}I_{t}\right]}{\mathbb{E}\left[\sum_{t=1}^{\tau}\{A_{t}-\tilde{\pi}(L_t)\}A_{t}I_{t}\right]},\]
where $\tilde{\pi}(L_t) = \tilde{\mathbb{E}}(A_{t}|L_{t},I_{t}=1) ={\alpha}_0+{\alpha}_1t+{\alpha}_2L_{t}$ is the fitted value from the population OLS regression of $A_t$ on $L_t$, $t$ and 1 within eligible individuals \supercite{robins1992}. The expression  of the population limit can be further rearranged. Considering the numerator:
\begin{flalign*}
    &\mathbb{E}\left[\sum^\tau_{t=1}\{A_{t}-\tilde{\pi}(L_t)\}Y_{t}I_t \right] \\
    &= \sum^\tau_{t=1}\mathbb{E}\left[\{A_{t}-\tilde{\pi}(L_t)\}Y_{t}I_t \right] \\
    &= \sum^\tau_{t=1}P(I_t=1)\mathbb{E}\left[\{A_{t}-\tilde{\pi}(L_t)\}Y_{t}|I_t=1 \right] + 0&\text{binary}\;I_t
\end{flalign*}\vspace{-1em}
\noindent 
The previous expectation equals:
\begin{flalign*}
    &\mathbb{E}\left[\{A_{t}-\tilde{\pi}(L_t)\}Y_{t}|I_t=1 \right] = \mathbb{E}\biggl(\mathbb{E}\left[\{A_{t}-\tilde{\pi}(L_t)\}Y_{t}|L_t,I_t=1 \right] \biggl|I_t=1\biggl)&\text{it.exp.}\\
    &= \mathbb{E}\biggl(\pi(L_t)\mathbb{E}\left[\{A_{t}-\tilde{\pi}(L_t)\}Y_{t}|A_t=1,L_t,I_t=1 \right] + \{1-\pi(L_t)\}\mathbb{E}\left[\{A_{t}-\tilde{\pi}(L_t)\}Y_{t}|A_t=0,L_t,I_t=1 \right]\biggl|I_t=1\biggl)&\text{binary}\;A_t\\
    &= \mathbb{E}\biggl(\pi(L_t)\mathbb{E}\left[\{1-\tilde{\pi}(L_t)\}Y_{t}|A_t=1,L_t,I_t=1 \right] + \{1-\pi(L_t)\}\mathbb{E}\left[\{0-\tilde{\pi}(L_t)\}Y_{t}|A_t=0,L_t,I_t=1 \right]\biggl|I_t=1\biggl) \\
    &= \mathbb{E}\biggl(\mathbb{E}\left[\pi(L_t)\{1-\tilde{\pi}(L_t)\}Y^1_{t}|A_t=1,L_t,I_t=1 \right] - \mathbb{E}\left[\{1-\pi(L_t)\}\tilde{\pi}(L_t)Y^0_{t}|A_t=0,L_t,I_t=1 \right]\biggl|I_t=1\biggl)&\text{cons.} \\
    &= \mathbb{E}\biggl(\mathbb{E}\left[\pi(L_t)\{1-\tilde{\pi}(L_t)\}Y^1_{t}|L_t,I_t=1 \right] - \mathbb{E}\left[\{1-\pi(L_t)\}\tilde{\pi}(L_t)Y^0_{t}|L_t,I_t=1 \right]\biggl|I_t=1\biggl)&\text{exch.}\\
    &= \mathbb{E}\biggl(\mathbb{E}\left[\pi(L_t)\{1-\tilde{\pi}(L_t)\}(Y^1_{t}-Y^0_{t})|L_t,I_t=1 \right] + \mathbb{E}\left[\{\pi(L_t)-\tilde{\pi}(L_t)\}Y^0_{t}|L_t,I_t=1 \right]\biggl|I_t=1\biggl)\\
    &= \mathbb{E}\left[\pi(L_t)\{1-\tilde{\pi}(L_t)\}(Y^1_{t}-Y^0_{t})|I_t=1 \right] + \mathbb{E}\left[\{\pi(L_t)-\tilde{\pi}(L_t)\}Y^0_{t}|I_t=1 \right].
\end{flalign*}
At the third equality we defined $\pi(L_t) = P(A_{t}=1|L_{t},I_{t}=1) = \mathbb{E}(A_{t}|L_{t},I_{t}=1)$. At the sixth equality, we added and subtracted $\mathbb{E}\left[\pi(L_t)\{1-\tilde{\pi}(L_t)\}Y^0_{t}|L_t,1 \right]$. The numerator can then be rewritten as:
\begin{flalign*}
        \sum^\tau_{t=1}P(I_t=1)\biggl(\mathbb{E}\left[\pi(L_t)\{1-\tilde{\pi}(L_t)\}(Y^1_{t}-Y^0_{t})|I_t=1 \right] + \mathbb{E}\left[\{\pi(L_t)-\tilde{\pi}(L_t)\}Y^0_{t}|I_t=1 \right]\biggl).
\end{flalign*}
As for the denominator:
\begin{flalign*}
    &\mathbb{E}\left[\sum^\tau_{t=1}\{A_{t}-\tilde{\pi}(L_t)\}A_tI_t \right] = \sum^\tau_{t=1}\mathbb{E}\left[\{A_{t}-\tilde{\pi}(L_t)\}A_tI_t \right] \\
    &= \sum^\tau_{t=1}P(I_t=1)\mathbb{E}\left[\{A_{t}-\tilde{\pi}(L_t)\}A_t|I_t=1 \right] &\text{binary}\;I_t \\
    &= \sum^\tau_{t=1}P(I_t=1)\mathbb{E}\biggl[\mathbb{E}\left[\{A_{t}-\tilde{\pi}(L_t)\}A_t|L_t,I_t=1 \right]\biggl|I_t=1\biggl]&\text{it.exp.}\\ 
    &= \sum^\tau_{t=1}P(I_t=1)\mathbb{E}\biggl[\pi(L_t)\mathbb{E}\left[\{1-\tilde{\pi}(L_t)\}\cdot1|A_t=1,L_t,I_t=1 \right] + 0 \; \biggl|I_t=1\biggl]&\text{binary}\;A_t\\ 
    &= \sum^\tau_{t=1}P(I_t=1)\mathbb{E}\biggl[\pi(L_t)\{1-\tilde{\pi}(L_t)\}\mathbb{E}\left[1|A_t=1,L_t,I_t=1 \right]\biggl|I_t=1\biggl]\\ 
    &= \sum^\tau_{t=1}P(I_t=1)\mathbb{E}\left[\pi(L_t)\{1-\tilde{\pi}(L_t)\}|I_t=1\right]
\end{flalign*}
\noindent from which the final expression of the asymptotic limit of the OLS estimator for $\psi$ is:
\begin{flalign*}
    \frac{\sum^\tau_{t=1}P(I_t=1)\mathbb{E}\left[\pi(L_t)\{1-\tilde{\pi}(L_t)\}(Y^1_{t}-Y^0_{t})|I_t=1 \right] }{\sum^\tau_{t=1}P(I_t=1)\mathbb{E}\left[\pi(L_t)\{1-\tilde{\pi}(L_t)\}|I_t=1\right]} + \frac{ \sum^\tau_{t=1}P(I_t=1)\mathbb{E}\left[\{\pi(L_t)-\tilde{\pi}(L_t)\}Y^0_{t}|I_t=1 \right]}{\sum^\tau_{t=1}P(I_t=1)\mathbb{E}\left[\pi(L_t)\{1-\tilde{\pi}(L_t)\}|I_t=1\right]}.
\end{flalign*}
This expression consists of two contributions. The first is a weighted average of the conditional contrast $\mathbb{E}\left[(Y^1_{t}-Y^0_{t})|L_t,I_t=1 \right]$, which is informative of the (conditional) treatment effect. The second is a weighted average of the difference $\pi(L_t)-\tilde{\pi}(L_t)$, which is generally non-zero unless either the linear outcome model is correctly specified or $\pi(L_t)$ is linear in $L_t$.{\supercite{robins1992}} This makes the estimand targeted by the OLS estimator a questionable summary of the treatment effect under model misspecification.

\subsection{Limit expression of the g-estimator}
\label{sec:asympt-gest}
Under the semiparametric model (3) of the main text, we can use the g-estimator to estimate $\psi$:
\[\hat{\psi}_{E}=\frac{\sum^n_{i=1}\sum_{t=1}^{\tau}\{A_{t,i}-\hat{\mathbb{E}}(A_{t,i}|L_{t,i},I_{t,i}=1)\}Y_{t,i}I_{t,i}}{\sum^n_{i=1}\sum_{t=1}^{\tau}\{A_{t,i}-\hat{\mathbb{E}}(A_{t,i}|L_{t,i},I_{t,i}=1)\}A_{t,i}I_{t,i}}\]
where $\hat{\mathbb{E}}(A_{t,i}|L_{t,i},I_{t,i}=1)$ is a consistent estimator of the propensity score $\mathbb{E}(A_{t,i}|L_{t,i},I_{t,i}=1)$. Note that the OLS estimator can be viewed as a special case of the g-estimator, where a linear propensity score model is postulated and fitted using least squares.
\\\\
The population limit of the g-estimator is:
\begin{align*}
\frac{\mathbb{E}\left[\sum^\tau_{t=1} \{A_{t}-\mathbb{E}(A_{t}|L_{t},I_{t}=1)\}Y_{t}I_t \right]}{\mathbb{E}\left[\sum^\tau_{t=1} \{A_{t}-\mathbb{E}(A_{t}|L_{t},I_{t}=1)\} A_tI_t\right]}.
\end{align*}
By proceeding similarly as for the OLS estimator, the population limit can be rewritten as:
\begin{align*}
\frac{\sum^\tau_{t=1}P(I_t=1)\mathbb{E}\left[Var(A_t|L_t,I_t=1)(Y^1_{t}-Y^0_{t}) |I_t=1\right]}{\sum^\tau_{t=1}P(I_t=1)\mathbb{E}\left[Var(A_{t}|L_t,I_t=1)|I_t=1 \right]}
\end{align*}
where $Var(A_{t}|L_t,I_t=1) = \pi(L_t)\{1-\pi(L_t)\}$. Expression (4) in the main text is then obtained by choosing $I_t$ as $1-A_{t-1}$. 

\subsection{Noncollapsible estimands}
In the simulation of Section \textit{Noncollapsible effect measures} of the main text, data are generated according to:
\begin{flalign*}
&L_t \; \sim \; N\biggl(0,A_{t-1} + \frac{1-A_{t-1}}{t}\biggl)\\
&A_t \; \sim \;  \text{Bernoulli}(0.2)\\
&Y_t \; \sim \;  \text{Bernoulli}\left\{expit( A_t + L_t)\right\}
\end{flalign*}
with $t = 1,\ldots,5$. The outcome was also generated as a continuous variable, according to:
\begin{flalign*}
Y_t \; \sim \;  N(A_t + L_t,1).
\end{flalign*}

\section{Model-free estimands for target trial emulation}
\subsection{Inspirations for the proposed estimands}
The proposed estimands are inspired by ideas from related fields. In the \emph{cluster randomized trials} literature, Kahan et al. \supercite{kahan2023estimands} discuss the \textit{participant-average treatment effect}, where every participant is given equal weight, and the \textit{cluster-average treatment effect}, where each cluster is weighted equally. These two effects respectively correspond to our $\psi_e$ (equation 6 in the main text) and $\psi_u$ (5). Furthermore, the calendar-time design finds analogies in the \emph{multicenter trials} literature:{\supercite{goldberg1990design}} each time point corresponds to a different center, with its associated (conditional) treatment effect, but with the additional complication that the same patient might belong to multiple centers. Alene et al.{\supercite{muluneh2025}} have proposed similar weighted estimands in this context.
The baseline-adjusted effect (7) is inspired by the \textit{meta-analysis} literature, which emphasizes defining causal inference with respect to a clear target population when transporting inferences from potentially heterogeneous studies \supercite{Bareinboim2016}. For example, Vo et al. \supercite{Tat-Thang2019} define an average treatment effect standardized to the baseline covariate distribution of one specific trial (e.g., trial $t=1$ in our proposal). Dahabreh et al. \supercite{Dahabreh2020} make a similar development, standardizing to the shared baseline covariate distribution across studies.
\\\\
A recent paper by Benz et al.{\supercite{benz2025statistical}} has some conceptual overlap, but their focus is on calendar-time-specific effects, while we consider effects aggregated across trials.

\subsection{An alternative motivation of the effects in a random trial}
\label{sec:effect_rt}
In a calendar-time setting, each trial comprises a group of eligible individuals sampled from the same superpopulation, which consists of all observations from all individuals present in the study between time 1 and $\tau$. For each individual, let $T$ be an artificial random variable representing a random visit index, following a discrete uniform distribution.
We can then define the estimand
\begin{align*}
    &\mathbb{E}(Y_{T}^1-Y_{T}^0|I=1),
\end{align*}
which is the counterfactual contrast at a randomly selected time $T$, conditional on eligible observations in the population. 
Given the uniformity of $T$, it is possible to rewrite the expression above as the eligibility weighted effect:
\begin{align*}
&\mathbb{E}(Y_{T}^1-Y_{T}^0|I=1)\\
&=\sum_{t=1}^{\tau} \mathbb{E}(Y^1_t-Y_t^0|I=1,T=t)P(T=t|I=1)&\text{it.exp.}\\
&=\sum_{t=1}^{\tau} \mathbb{E}(Y^1_t-Y_t^0|I=1,T=t)P(I=1|T=t)\frac{P(T=t)}{P(I=1)}&\text{Bayes' rule}\\
&=\sum_{t=1}^{\tau} \mathbb{E}(Y^1_t-Y_t^0|I=1,T=t)\frac{P(I=1|T=t)P(T=t)}{\sum_{j=1}^{\tau}P(I=1|T=j)P(T=j)}&\text{tot. prob.}\\
&=\sum_{t=1}^{\tau} \mathbb{E}(Y^1_t-Y_t^0|I_t=1)\frac{P(I_t=1)P(T=t)}{\sum_{j=1}^{\tau}P(I_j=1)P(T=j)}&P(I=1|T=t)=P(I_t=1)\\
&=\sum_{t=1}^{\tau} \mathbb{E}(Y^1_t-Y_t^0|I_t=1)\frac{P(I_t=1)}{\sum_{j=1}^{\tau} P(I_j=1)}&\text{uniformly distributed } F(T)
\end{align*}
which is the eligibility weighted effect. Above and in the following sections, \textit{it.exp.} means the law of iterated expectation, while \textit{tot. prob.} means the law of total probability. By further assuming the same probability of being eligible across trials, we derive the uniformly weighted effect:
\begin{align*}
\psi_u 
= & \; \frac{1}{\tau}\sum_{t=1}^{\tau} \mathbb{E}(Y^1_t-Y_t^0|I_t=1).
\end{align*}

\subsection{Motivation for $\psi_u$ and $\psi_e$ in visit-time designs}
While $\psi_u$ and $\psi_e$ were primarily conceived in the calendar-time setting, they remain well-defined and potentially informative in visit-time designs.
Estimand $\psi_e$ can be of interest when the research question concerns the average treatment effect across all eligible patient-time units, regardless of time since the anchor event. For instance, in an ICU study where patients are followed across several visits, $\psi_e$ reflects the effect for a patient selected at a random point during their ICU stay, providing an informative summary for a clinician concerned about the average benefit of treatment irrespective of how long patients have been admitted.
Estimand $\psi_u$ may instead  be useful when the researcher wishes to give equal weight to each time point, regardless of how many patients remain eligible, with the caveat that patients who remain in care longer are implicitly up-weighted. For example, this might be of interest if one wishes to study the effect of a treatment (e.g., early mobilization) across all ICU days treated as equally relevant, answering: \textit{"What is the effect of early mobilization on a randomly chosen ICU day?"}.
Consider that in visit-time settings with a strong selective trend, the target population implied by $\psi_u$ might become progressively less representative of the clinically relevant population, as it will increasingly reflect the effect within narrower populations at later time points (e.g., healthier patients). In these cases, $\psi_b$ is typically preferable.

\subsection{Interpretation of the estimands under effect modification within strata of $L_t$}
Here we illustrate the interpretation of the three proposed estimands when the treatment effect is modified by a time-varying covariate $L_t$ which is predictive of the outcome.
Consider a simple example with 10 eligible patients at $t=1$, and 5 eligible patients at $t=2$. The binary covariate $L_t \in \{0,1\}$ indicates whether a patient is healthy $(L_t=1)$ or not $(L_t=0)$. The treatment effect is modified by $L_t$, so that:
\begin{flalign*}
    &\theta_t(1) = \mathbb{E}(Y^1_t-Y_t^0|L_t=1,I_t=1) = 2\\
    &\theta_t(0) = \mathbb{E}(Y^1_t-Y_t^0|L_t=0,I_t=1) = 1
\end{flalign*}
representing the fact that treatment has a stronger effect among healthier patients. For simplicity, we assume that the treatment effect within subgroups of $L_t$ does not change over time. We consider the population of eligible patients to be distributed as follows:
    \vspace{20pt}
\begin{center}
\begin{tabular}{lccc}
    \toprule
     Trial & Healthy patients ($L_t=1$) & Unhealthy patients ($L_t=0$) & Total \\
    \midrule
    $t=1$ & 5 (50\%) & 5 (50\%) & 10 \\
    $t=2$ & 4 (80\%) & 1 (20\%) & 5  \\
    \bottomrule
\end{tabular}
\end{center}    
\vspace{20pt}
While the eligible patients at baseline are equally distributed among healthy and unhealthy, trial 2 contains a larger percentage of healthy patients, representing the expected selective trend in visit-time trials. Thus, the time-specific effects would be:
\begin{flalign*}
    &\theta_1 = 2 \times 0.5 + 1 \times0.5 = 1.5\\
    &\theta_2 =  2 \times 0.8 + 1 \times0.2 = 1.8
\end{flalign*}
where the difference is due to the shift of the covariate distribution, not to the treatment becoming more effective. The three estimands are:
\begin{flalign*}
    &\psi_u = \frac{1}{2} (1.5+1.8) = 1.65\\
    &\psi_e = \frac{10}{15}\times1.5 + \frac{5}{15}\times1.8 = 1.6\\
    &\psi_b = \frac{1}{2}\left(0.5\times2+0.5\times1\right) + \frac{1}{2}\left(0.5\times2+0.5\times1\right)  = 1.5
\end{flalign*}
The three estimands answer different questions: $\psi_u$ asks \textit{"What is the expected treatment effect in a randomly selected trial?"}, giving equal weight to both trials and thus highlighting the healthier population at $t=2$ (in general, this can be relevant when the research question concerns the average effect across time points that are treated as equally relevant); $\psi_e$ asks \textit{"What is the expected treatment effect for a randomly selected patient-time?"}, giving more weight to the larger trial at $t=1$; and $\psi_b$ asks \textit{"What is the expected treatment effect for a patient from the baseline population?"}, regardless of how the patients' characteristics evolve with time. The choice among them should be guided by the target population most relevant to the clinical question. A clinician interested in the effect for the typical patient currently in care, regardless of time since admission, would prefer $\psi_e$; one interested in the effect for the average trial would consider $\psi_u$; one interested in the effect for the baseline population would prefer $\psi_b$.

\subsection{Identification of Causal Estimands}
\label{sec:appendix-identification}
Here we provide the derivations of the estimators for the proposed estimands. For this purpose, we rely on the following assumptions:
\begin{itemize}
    \item \textbf{Sequential exchangeability}
    \begin{flalign*}
    A_t\indep Y^{a_t}_{t}|\overline{W}_{t},I_{t}=1 \qquad\forall \; t,a_t&&
    \end{flalign*}    
    \item \textbf{Consistency}
    \begin{flalign*}
    A_t = a_t \implies Y_{t} = Y^{a_t}_{t} \qquad\forall \; t,a_t&&
    \end{flalign*}      
    \item \textbf{Positivity of treatment}
    \begin{flalign*}
        P(A_{t}=a_t|\overline{W}_{t},I_{t}=1)>0 \qquad \mbox{w.p.1} \qquad \forall \; t,a_t&&
    \end{flalign*}      
    \item \textbf{Positivity of eligibility}
    \begin{itemize}
        \item For estimating $\psi_u$: $P(I_{t}=1)>0 \qquad\forall \; t$  
        \item For estimating $\psi_e$: $P(I_{t}=1)>0$ for at least one $t$ 
        \item For estimating $\psi_b$: $P(I_{t}=1|L_1)>0 \qquad \mbox{ w.p.1} \qquad\forall \; t >1$
    \end{itemize}
\end{itemize}
As we can see, besides the standard positivity assumptions for time-varying treatments, the proposed estimators rely on further positivity assumptions regarding eligibility. The estimation of $\psi_u$ demands eligible patients in all trials, whereas the estimation of $\psi_e$ merely demands the presence of eligible patients in at least one trial. For the estimation of $\psi_b$, we need the stronger requirement that no individuals in the baseline population are prevented from reaching any trial $t>1$. This condition is more restrictive, and related to the \textit{positivity of the probability of participation in the trials} assumption in the transportability literature, such as the positivity assumption \textit{b} in Vo et al.\supercite{Tat-Thang2019} or assumption \textit{B5} in Dahabreh et al.\supercite{Dahabreh2020} Note that we conceptualize visit-time trials as a progressively narrowing closed cohort, so that all patients are eligible at baseline. This is a peculiarity of the visit-time design (see Appendix S1.1) and it implies a restriction on eligibility criteria that researchers wishing to implement $\psi_b$ must account for, since any patient appearing in a later trial must already be present at baseline. For example, baseline eligibility cannot depend on $L_1$, since this would prevent patients with certain levels of $L_1$ from being eligible at $t=1$. In contrast, if one allowed patients to be ineligible at baseline, the required condition would be $P(I_t=1\mid I_1=1,L_1)>0$ w.p.1 $\;\forall\, t$, and the baseline-adjusted effect would need to condition on $I_1=1$ in the outer expectation: $\psi_b = \frac{1}{\tau}\sum^\tau_{t=1}\mathbb{E}\left\{\mathbb{E}\left(Y^1_{t}-Y^0_{t}\mid I_t=1,L_1\right)\mid I_1=1\right\}$, thus standardizing to the distribution of $L_1$ among eligible patients at baseline.

\subsubsection{Baseline-adjusted effect}\label{sec:id-baseline}
The IPW and G-computation estimators for the baseline-adjusted effect $\psi_b$ are derived here.\\\\
\noindent
\textbf{IPW estimator}\\
Consider the baseline-adjusted effect (7). Assuming the identifiability conditions stated above, it is possible to work out the time-specific contribution:\\
\begin{flalign*}
\mathbb{E}\left\{\mathbb{E}(Y^1_{t}-Y^0_{t}|L_1,I_t=1)\right\} \; = \;  \mathbb{E}\left\{\underbrace{\mathbb{E}(Y^1_{t}|L_1,I_t=1)}_{\boxed{A}} -\underbrace{\mathbb{E}(Y^0_{t}|L_1,I_t=1)}_{\boxed{B}} \right\}.
\end{flalign*}
We can rewrite expectation $\boxed{A}$:
\begin{flalign*}
\boxed{A} &= \mathbb{E}(Y^1_{t}|L_1,I_t=1) \\
&= \mathbb{E} \left\{\mathbb{E}(Y^1_{t}|\overline{W}_t,I_t=1)|L_1,I_t=1 \right\} &\text{it.exp.}\\
&= \mathbb{E} \left\{ \frac{P(A_t=1|\overline{W}_t,I_t=1)}{P(A_t=1|\overline{W}_t,I_t=1)}\mathbb{E}(Y^1_{t}|\overline{W}_t,I_t=1)\biggl|L_1,I_t=1 \right\}&\text{positivity of }A_t &&\\
&= \mathbb{E} \left\{ \frac{\mathbb{E}(A_t|\overline{W}_t,I_t=1)}{P(A_t=1|\overline{W}_t,I_t=1)}\mathbb{E}(Y^1_{t}|\overline{W}_t,I_t=1)\biggl|L_1,I_t=1 \right\}&\text{binary $A_t$} &&\\
&= \mathbb{E} \left\{ \frac{\mathbb{E}(A_tY^1_{t}|\overline{W}_t,I_t=1)}{P(A_t=1|\overline{W}_t,I_t=1)}\biggl|L_1,I_t=1 \right\}&\text{$A_t \perp\!\!\!\perp Y^1_{t}|\overline{W}_t,I_t=1$} &&\\
&= \mathbb{E} \left\{ \frac{\mathbb{E}(A_tY_{t}|\overline{W}_t,I_t=1)}{P(A_t=1|\overline{W}_t,I_t=1)}\biggl|L_1,I_t=1 \right\}&\text{consistency} &&\\
&= \mathbb{E} \left\{ \frac{A_tY_{t}}{P(A_t=1|\overline{W}_t,I_t=1)}\biggl|L_1,I_t=1 
\right\}.&\text{it.exp.}
\end{flalign*} 
By proceeding similarly, we can rewrite $\boxed{B}$ as:
\begin{flalign*}
\boxed{B} &= \mathbb{E}(Y^0_{t}|L_1,I_t=1)&&\\
&=\mathbb{E} \left\{\mathbb{E}(Y^0_{t}|\overline{W}_t,I_t=1)|L_1,I_t=1 \right\}\\
&= \mathbb{E} \left[\frac{(1-A_t)Y_{t}}{\{1- P(A_t=1|\overline{W}_t,I_t=1)\}}\biggl|L_1,I_t=1 \right].
\end{flalign*} 
\noindent
The initial causal effect can be expressed as:
\vspace{-0.2cm}
\begin{flalign*}
    &\mathbb{E}\left\{\mathbb{E}(Y^1_{t}-Y^0_{t}|L_1,I_t=1)\right\}&&\\
    &= \mathbb{E}\left(\underbrace{\mathbb{E} \left\{ \frac{A_tY_{t}}{P(A_t=1|\overline{W}_t,I_t=1)}\biggl|L_1,I_t=1 \right\}}_{\boxed{C}} -  \underbrace{\mathbb{E} \left[\frac{(1-A_t)Y_{t}}{\{1- P(A_t=1|\overline{W}_t,I_t=1)\}}\biggl|L_1,I_t=1 \right]}_{\boxed{D}} \right),
\end{flalign*}
where the two expectations can be further rearranged. Considering expectation $\boxed{C}$:
\begin{flalign}
    &\mathbb{E}\left[\mathbb{E} \left\{ \frac{A_tY_{t}}{P(A_t=1|\overline{W}_t,I_t=1)}\biggl|L_1,I_t=1 \right\} \right]
    = \int \mathbb{E} \left\{ \frac{A_tY_{t}}{P(A_t=1|\overline{W}_t,I_t=1)}\biggl|L_1,I_t=1 \right\} f(l_1) dl_1&&\nonumber \\
    & = \int \mathbb{E} \left\{ \frac{A_tY_{t}}{P(A_t=1|\overline{W}_t,I_t=1)}\biggl|L_1,I_t=1 \right\} f(l_1) \frac{f(l_1|I_t=1)}{f(l_1|I_t=1)} dl_1 &&\nonumber \\
    & = \int \mathbb{E} \left\{ \frac{A_tY_{t}}{P(A_t=1|\overline{W}_t,I_t=1)}\biggl|L_1,I_t=1 \right\} f(l_1) \frac{f(l_1|I_t=1)P(I_t=1)}{P(I_t=1|l_1)f(l_1)} dl_1 \qquad\text{Bayes' theorem, }P(I_t=1,l_1)>0 &&\nonumber \\
    & = \int \mathbb{E} \left\{ \frac{A_tY_{t}}{P(A_t=1|\overline{W}_t,I_t=1)} \frac{P(I_t=1)}{P(I_t=1|L_1)}\biggl|L_1,I_t=1 \right\}f(l_1|I_t=1) dl_1 &&\nonumber \\
    & = \mathbb{E} \left\{ \frac{A_tY_{t}}{P(A_t=1|\overline{W}_t,I_t=1)} \frac{P(I_t=1)}{P(I_t=1|L_1)}\biggl|I_t=1 \right\}. \qquad\qquad\qquad\qquad\qquad\qquad\quad\text{it.exp.}&&\label{expect-1}\nonumber
\end{flalign}
The conditional expectation can be rewritten as:
\begin{flalign*}
    \nonumber&\nonumber = \mathbb{E} \left\{ \frac{A_tY_{t}}{P(A_t=1|\overline{W}_t,I_t=1)} \frac{P(I_t=1)}{P(I_t=1|L_1)}|I_t=1 \right\} &&\\\nonumber
    & = \int  \frac{a_ty_{t}}{P(A_t=1|\overline{w}_t,I_t=1)} \frac{P(I_t=1)}{P(I_t=1|l_1)}f(a_t,y_{t},l_1,\overline{w}_t|I_t=1) da_tdy_{t}dl_1d\overline{w}_t &&\\\nonumber
    & = \int  \frac{a_ty_{t}}{P(A_t=1|\overline{w}_t,I_t=1)} \frac{P(I_t=1)}{P(I_t=1|l_1)}\frac{f(a_t,y_{t},l_1,\overline{w}_t,I_t=1)}{P(I_t=1)} da_tdy_{t}dl_1d\overline{w}_t &&\text{Bayes' theorem}\\\nonumber
    & = \int  \frac{a_ty_{t}i_t}{P(A_t=1|\overline{w}_t,I_t=1)} \frac{P(I_t=1)}{P(I_t=1|l_1)}\frac{f(a_t,y_{t},l_1,\overline{w}_t,i)}{P(I_t=1)} da_tdy_{t}dl_1d\overline{w}_t &&\text{binary }I_t\\\nonumber
    &= \frac{1}{P(I_t=1)}\mathbb{E} \left\{ \frac{A_tY_{t}I_t}{P(A_t=1|\overline{W}_t,I_t=1)} \frac{P(I_t=1)}{P(I_t=1|L_1)} \right\} \\
    & = \mathbb{E} \left\{\frac{1}{P(I_t=1|L_1)} \frac{A_tY_{t}I_t}{P(A_t=1|\overline{W}_t,I_t=1)} \right\}.
\end{flalign*}
By proceeding similarly for the second expectation $\boxed{D}$, we obtain:
\begin{flalign*}
    &\mathbb{E}\left[\mathbb{E} \left\{\frac{(1-A_t)Y_{t}}{\{1- P(A_t=1|\overline{W}_t,I_t=1)\}}\biggl|L_1,I_t=1 \right\} \right]&&\\
    & = \mathbb{E} \left[ \frac{(1-A_t)Y_{t}}{\{1-P(A_t=1|\overline{W}_t,I_t=1)\}}\frac{P(I_t=1)}{P(I_t=1|L_1)}\biggl|I_t=1 \right]&&\\
    & = \mathbb{E} \left[\frac{1}{P(I_t=1|L_1)} \frac{(1-A_t)Y_{t}I_t}{\{1-P(A_t=1|\overline{W}_t,I_t=1)\}}\right].
\end{flalign*}
\noindent
Note that integrals can be replaced with sums when $L_1$ is discrete. Thus, by taking the average over time points, the baseline-adjusted effect reduces to the difference of observables:
\begin{equation*}
    \psi_b = \frac{1}{\tau}\sum^{\tau}_{t=1}\mathbb{E}\left(\frac{1}{P(I_t=1|L_1)} \biggl[ \frac{A_tY_{t}I_t}{P(A_t=1|\overline{W}_t,I_t=1)} - \frac{(1-A_t)Y_{t}I_t}{\{1-P(A_t=1|\overline{W}_t,I_t=1)\}}\biggl]\right).
\end{equation*}
This expression can be estimated as:
\begin{equation*}
    \hat{\psi}_{b-ipw} =\frac{1}{\tau}\frac{1}{n}\sum^{\tau}_{t=1}\sum^n_{i=1} \frac{1}{\hat{P}(I_{t,i}=1|L_{1,i})}  \biggl[\frac{A_{t,i}Y_{t,i}I_{t,i}}{\hat{P}(A_{t,i}=1|\overline{W}_{t,i},I_{t,i}=1)} - \frac{(1-A_{t,i})Y_{t,i}I_{t,i}}{\{1-\hat{P}(A_{t,i}=1|\overline{W}_{t,i},I_{t,i}=1)\}}\biggl]
\end{equation*}
\textbf{G-computation estimator}\\
Starting from the time-specific contribution for the baseline-adjusted effect, we can write:
\begin{flalign*}
    &\mathbb{E}\left\{\mathbb{E}(Y^1_t-Y^0_t|L_1,I_t=1)\right\}&&\\
    &= \mathbb{E}\left\{\mathbb{E}(Y^1_t|L_1,I_t=1) - \mathbb{E}(Y^0_t|L_1,I_t=1) \right\}&\text{lin.}&&\\
    &= \mathbb{E}\left[ \mathbb{E}\left\{\mathbb{E}(Y^1_t|I_t=1,\overline{W}_t)|L_1,I_t=1\right\} - \mathbb{E}\left\{\mathbb{E}(Y^0_t|I_t=1,\overline{W}_t)|L_1,I_t=1\right\} \right]&\text{it.exp.}&&\\
    &= \mathbb{E}\left[ \mathbb{E}\left\{\mathbb{E}(Y_t|A_t=1,I_t=1,\overline{W}_t)|L_1,I_t=1\right\} - \mathbb{E}\left\{\mathbb{E}(Y_t|A_t=0,I_t=1,\overline{W}_t)|L_1,I_t=1\right\} \right]&\text{cons.}&&
\end{flalign*}
which can be estimated as:
\begin{flalign*}
    \hat{\psi}_{b-gcomp} =\frac{1}{\tau}\frac{1}{n}\sum^{\tau}_{t=1}\sum^n_{i=1}\left[ \hat{\mathbb{E}}\left\{\hat{\mathbb{E}}(Y_{t}|A_t=1,I_t=1,\overline{W}_t)|L_1,I_t=1\right\} - \hat{\mathbb{E}}\left\{\hat{\mathbb{E}}(Y_{t}|A_t=0,I_t=1,\overline{W}_t)|L_1,I_t=1\right\} \right].
\end{flalign*}
\noindent
Note that in the current implementation of the estimators for the baseline-adjusted effect, some degree of model misspecification is to be expected. In fact, in the G-computation estimator of $\psi_b$, the nested outcome models condition on different sets of covariates, which may not necessarily be compatible. A similar issue can arise for the IPW estimator of $\psi_b$, where $P(I_{t,i}=1|L_{1,i})$ and $P(A_{t,i}=1|\overline{W}_{t,i},I_{t,i}=1)$ target the same quantity using different conditioning sets, considering that eligibility is defined as $I_t = 1-A_{t-1}$. As a result, both models cannot generally be correctly specified at the same time. 
However, our framework does not depend on any particular parametric specification of the nuisance functions. As we explain in the Discussion of the main text, arbitrarily complex parametric or nonparametric models can be employed to reduce the risk of misspecification, without altering the interpretation of the target estimand. 

\subsubsection{Uniformly weighted effect in a random trial}\label{sec:id-uniform}
Here we derive the IPW and G-computation estimators for the uniformly weighted effect.\\
\textbf{IPW}\\
Consider the time-specific contribution of the effect (5):
\begin{align*}
\mathbb{E}(Y^1_t-Y_t^0|I_t=1) = \biggl\{\underbrace{\mathbb{E}(Y^1_t|I_t=1)}_{\boxed{A}}-\underbrace{\mathbb{E}(Y_t^0|I_t=1)}_{\boxed{B}}\biggl\}.
\end{align*}
Then
\begin{flalign*}
\boxed{A} &= \mathbb{E}(Y_t^1|I_t=1)\\
&=\mathbb{E} \left\{\mathbb{E}(Y^1_t|\overline{W}_t,I_t=1)|I_t=1 \right\}&\text{it.exp.}\\
&=\mathbb{E} \left\{\frac{P(A_t=1|\overline{W}_t,I_t=1)}{P(A_t=1|\overline{W}_t,I_t=1)}\mathbb{E}(Y^1_t|\overline{W}_t,I_t=1)\biggl|I_t=1 \right\} &\text{pos.} \\
&=\mathbb{E} \left\{\frac{\mathbb{E}(A_t|\overline{W}_t,I_t=1)}{P(A_t=1|\overline{W}_t,I_t=1)}\mathbb{E}(Y^1_t|\overline{W}_t,I_t=1)\biggl|I_t=1\right\} &\text{binary } A_t\\
&=\mathbb{E} \left\{\frac{\mathbb{E}(A_t Y^1_t|\overline{W}_t,I_t=1)}{P(A_t=1|\overline{W}_t,I_t=1)}\biggl|I_t=1\right\} &A_t \perp\!\!\!\perp Y^1_t|\overline{W}_t,I_t=1\\
&=\mathbb{E} \left\{\frac{\mathbb{E}(A_t Y_t|\overline{W}_t,I_t=1)}{P(A_t=1|\overline{W}_t,I_t=1)}\biggl|I_t=1\right\} &\text{cons.}\\
&=\mathbb{E} \left\{\frac{A_t Y_t}{P(A_t=1|\overline{W}_t,I_t=1)}\biggl|I_t=1\right\} &\text{it.exp.}\\
&= \frac{1}{P(I_t = 1)} \mathbb{E}\biggl\{\frac{A_t Y_t I_t}{P(A_t=1|\overline{W}_t,I_t=1)}\biggl\} &
\end{flalign*}
and, by proceeding similarly for expectation $\boxed{B}$:
\begin{flalign*}
\boxed{B} &= \mathbb{E}(Y_t^0|I_t=1)&&\\
&=\mathbb{E} \left[\frac{(1-A_t) Y_t}{\{1-P(A_t=1|\overline{W}_t,I_t=1)\}}\biggl|I_t=1\right]\\
&= \frac{1}{P(I_t = 1)} \mathbb{E}\left[\frac{(1-A_t) Y_t I_t}{\{1-P(A_t=1|\overline{W}_t,I_t=1)\}}\right].
\end{flalign*} 
By taking the average across time points, the uniformly weighted effect can thus be rewritten as the difference of observables:
\begin{flalign*}
\psi_u &=  \frac{1}{\tau}\sum_{t=1}^{\tau}\frac{1}{P(I_t=1)}\mathbb{E}\biggl[\frac{A_t Y_t I_t}{P(A_t=1|\overline{W}_t,I_t=1)} - \frac{(1 - A_t) Y_t I_t}{\{1 - P(A_t=1|\overline{W}_t,I_t=1)\}} \biggl].
\end{flalign*}
This expression can be estimated as:
\begin{flalign*}
\hat{\psi}_{u-ipw} = \frac{1}{\tau}\frac{1}{n}\sum_{t=1}^{\tau}\sum_{i=1}^{n}\frac{1}{\hat{P}(I_{t} = 1)}\biggl[\frac{A_{t,i} Y_{t,i} I_{t,i}}{\hat{P}(A_{t,i}=1|\overline{W}_{t,i},I_{t,i} = 1)} - \frac{(1 - A_{t,i}) Y_{t,i} I_{t,i}}{\{1 - \hat{P}(A_{t,i}=1|\overline{W}_{t,i},I_{t,i} = 1)\}}\biggl].
\end{flalign*}
\textbf{G-computation:}\\
As for the G-computation estimator, the time-specific contrast can be rewritten as:
\begin{flalign*}
&\mathbb{E}(Y^1_t-Y_t^0|I_t=1) = \frac{1}{\tau}\sum_{t=1}^{\tau}\mathbb{E}(Y^1_t-Y_t^0|I_t=1)\\
&= \frac{1}{\tau}\sum_{t=1}^{\tau}\mathbb{E}(Y^1_t|I_t=1)-\mathbb{E}(Y_t^0|I_t=1)&\text{linearity}\\
&= \frac{1}{\tau}\sum_{t=1}^{\tau}\mathbb{E}\biggl[\mathbb{E}(Y^1_t|\overline{W}_{t},I_t=1)\biggl|I_t=1\biggl]-\mathbb{E}\biggl[\mathbb{E}(Y_t^0|\overline{W}_{t},I_t=1)\biggl|I_t=1\biggl]&\text{it.exp.}\\
&= \frac{1}{\tau}\sum_{t=1}^{\tau}\mathbb{E}\biggl[\mathbb{E}(Y_t|A_t = 1,\overline{W}_{t},I_t=1)\biggl|I_t=1\biggl]-\mathbb{E}\biggl[\mathbb{E}(Y_t|A_t = 0,\overline{W}_{t},I_t=1)\biggl|I_t=1\biggl]&\text{cons., exch.}\\
&= \frac{1}{\tau}\sum_{t=1}^{\tau}\mathbb{E}\biggl[\mathbb{E}(Y_t|A_t = 1,\overline{W}_{t},I_t=1) - \mathbb{E}(Y_t|A_t = 0,\overline{W}_{t},I_t=1)\biggl|I_t=1\biggl]
\end{flalign*}
and the corresponding estimator:
    \begin{flalign*}
\hat{\psi}_{u-gcomp}= \frac{1}{\tau}\frac{1}{n}\sum_{t=1}^{\tau}\sum_{i=1}^{n}\frac{I_{t,i}}{\hat{P}(I_{t} = 1)}\biggl\{\hat{\mathbb{E}}(Y_{t,i}|A_{t,i} = 1,\overline{W}_{t,i},I_{t,i}=1) - \hat{\mathbb{E}}(Y_{t,i}|A_{t,i} = 0,\overline{W}_{t,i},I_{t,i}=1)\biggl\}
\end{flalign*}

\subsubsection{Eligibility weighted effect in a random trial}\label{sec:id-elig}
Following similar steps as we previously did for the uniformly weighted effect in a random trial, we derive the IPW and G-computation estimators for the eligibility weighted effect (6).\\
\textbf{IPW:}
\begin{align*}
\sum_{t=1}^{\tau}\mathbb{E}(Y^1_t-Y_t^0|I_t=1)\frac{P(I_t=1)}{\sum_{t=1}^{\tau} P(I_t=1)}
\end{align*}
which can be estimated as:
\begin{align*}
\hat{\psi}_{e-ipw} &= \frac{1}{n}\sum_{i=1}^{n}\sum_{t=1}^{\tau}\frac{1}{\hat{P}(I_{t} = 1)}\biggl[\frac{A_{t,i} Y_{t,i} I_{t,i}}{\hat{P}(A_{t,i}=1|\overline{W}_{t,i},I_{t,i}=1)} - \frac{(1 - A_{t,i}) Y_{t,i} I_{t,i}}{\{1 - \hat{P}(A_{t,i}=1|\overline{W}_{t,i},I_{t,i}=1)\}}\biggl]\frac{\hat{P}(I_{t,i} = 1)}{\sum_{t=1}^{\tau} \hat{P}(I_{t,i} = 1)}\\
&= \frac{1}{n}\sum_{i=1}^{n}\sum_{t=1}^{\tau}\frac{1}{\{\sum_{t=1}^{\tau} \hat{P}(I_{t} = 1)\}}\biggl[\frac{A_{t,i} Y_{t,i} I_{t,i}}{\hat{P}(A_{t,i}=1|\overline{W}_{t,i},I_{t,i}=1)} - \frac{(1 - A_{t,i}) Y_{t,i} I_{t,i}}{\{1 - \hat{P}(A_{t,i}=1|\overline{W}_{t,i},I_{t,i}=1)\}}\biggl]
\end{align*}
\vspace{1cm}
\textbf{G-computation:}\\
As before, this can be estimated as:
\begin{align*}
\hat{\psi}_{e-gcomp} = \frac{1}{n}\sum_{t=1}^{\tau}\sum_{i=1}^{n}\frac{I_{t,i}}{\{\sum_{t=1}^{\tau} \hat{P}(I_{t}=1)\}}\biggl\{\hat{\mathbb{E}}(Y_{t,i}|A_{t,i} = 1,\overline{W}_{t,i},I_{t,i}=1) - \hat{\mathbb{E}}(Y_{t,i}|A_{t,i} = 0,\overline{W}_{t,i},I_{t,i}=1)\biggl\}
\end{align*}

\section{Simulation study}
\label{appendix:sim}

In the simulation studies, data are generated under both visit-time and calendar-time trial designs. In the calendar-time design, individuals may exit the study at any time point $t$ with a fixed probability. When a participant is discharged, a new individual enters the study, maintaining a constant number of participants across calendar dates. Participants are eligible for a trial when they have not previously received treatment (i.e., $I_t=1-A_{t-1}$). Newly entered individuals are assumed to be treatment-naive, ensuring that $I_1 = 1$. We generate data for two time points, structured according to the causal diagram shown in Figure \ref{fig:dag}.
\begin{figure}[H]
\centering
\scalebox{0.75}{
\begin{tikzpicture}[
    >=Stealth 
]
    \node (L1) at (0,10) {$L_1$};
    \node (A1) at (0,8) {$A_1$};
    \node (Y1) at (1.3,6) {$Y_1$};

    \node (L2) at (2,10) {$L_2$};
    \node (A2) at (2,8) {$A_2$};
    \node (Y2) at (3.3,6) {$Y_2$};



    \draw[->] (L1) -- (Y1);
    \draw[->] (L1) -- (A1);
    \draw[->] (L1) -- (L2);
    \draw[->] (A1) -- (Y1);
    \draw[->] (A1) -- (L2);
    \draw[->] (A1) -- (A2);
    \draw[->] (Y1) -- (A2);
    \draw[->] (Y1) -- (Y2);
    \draw[->] (L2) -- (Y2);
    \draw[->] (L2) -- (A2);
    \draw[->] (A2) -- (Y2);

\end{tikzpicture}
}
\caption[DAG of the data-generating process used in the simulations]{The DAG represents a data-generating process we use to artificially generate the data.}
\label{fig:dag}\vspace{0.5cm}
\RaggedRight\textit{Alt text:} Directed acyclic graph of the data-generating process across two time points. Top row: L1 (left), L2 (right). Middle row: A1 (left), A2 (right). Bottom row: Y1 (left), Y2 (right).

\end{figure}

\begin{figure}[t]
\centering
\begin{subfigure}{0.37\textwidth}
    \includegraphics[width=\textwidth]{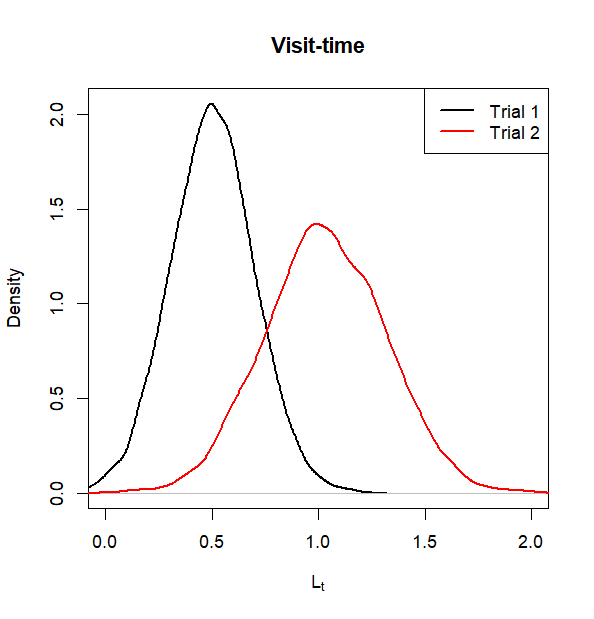}
    \caption{}
\end{subfigure}
\hspace{1cm}
\begin{subfigure}{0.37\textwidth}
    \includegraphics[width=\textwidth]{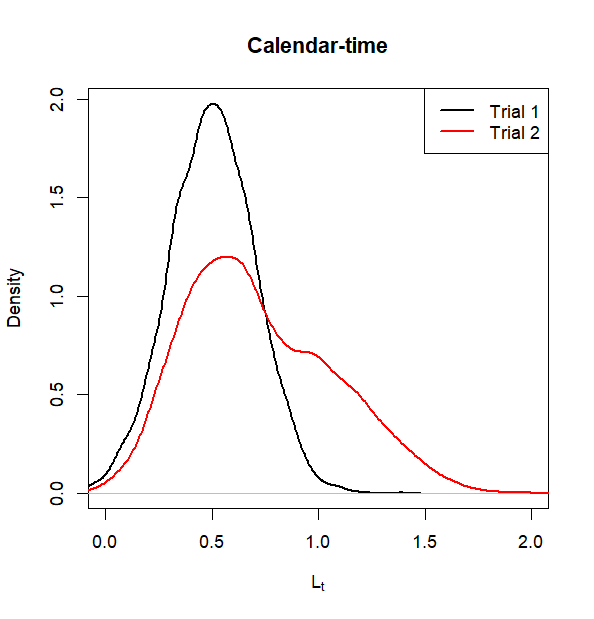}
    \caption{}
\end{subfigure}

\caption[Distributions of $L_t$ among eligible patients at trials 1 and 2]{Average distributions of $L_t$ among eligible patients at trial 1 (black) 
    and trial 2 (red), across 1000 simulated datasets of size 1000. 
    Left: visit-time design. Right: calendar-time design.}
    \label{fig_S2}\vspace{0.5cm}
\RaggedRight\textit{Alt text:} Two density plots comparing the distribution of the time-varying covariate $L_t$ among eligible patients at trial 1 (black curve) versus trial 2 (red curve), averaged across 1000 simulated datasets of size 1000. Left panel (visit-time design): black curve peaks sharply around 0.5, while red is shifted right, peaking lower and broader around $L_t$ = 1. Right panel (calendar-time design): the black curve is nearly identical to the left panel, but the red curve is flatter and bimodal, with a main peak near 0.5 and a smaller secondary bump near 0.9, overlapping more with the black curve than in the visit-time design.
\end{figure}
\noindent
The time-dependent variables are sequentially generated as:
\begin{align*}
&L_t \; \sim \; N(\alpha_0 + \alpha_1 L_{t-1}  + \alpha_2 A_{t-1}, 0.2)\\
&A_t \; \sim \;  A_{t-1} + (1-A_{t-1}) \cdot  \text{Bernoulli}\{ p = expit(\beta_0 + \beta_1 L_{t} +  \beta_2 Y_{t-1})\}\\
&Y_t \; \sim \;  N(\gamma_0 + \gamma_1 A_t + \gamma_2 L_t + \gamma_3 Y_{t-1}, 0.5)
\end{align*}
for every $t$, both in the visit-time and calendar-time designs; in the calendar-time design, new patients are allowed to enter the study at any $t$, while in the visit-time design no new patients enter after $t=1$. Here, at each patient’s first occurrence, all variables indexed $t-1$ are defined as 0. Furthermore, $A_t$ is generated so that once a patient initiates treatment, they remain treated at all subsequent time points, and so are no longer eligible. Note that the treatment effect corresponds to $\gamma_1$. 
Figure \ref{fig_S2} shows the plots of the average distributions of $L_1$ and $L_2$ among eligible patients for the two designs across 1000 simulations. The data are generated so that for visit-time data, the characteristics of the eligible population change considerably between the two trials; for calendar-time data, the change in the distribution of the covariate is smaller. The distributions are the same for either setting of the simulation study.


\subsection{Time-dependent treatment effect, binary outcome}
\label{additional-sim}

Here we present the results of an additional simulation study with binary outcome, where conventional approaches often rely on noncollapsible estimands. We performed two different simulation designs (see Table~{\ref{table:dgm_setting3}}). For the calendar-time data, we generated the outcome as:
\begin{align}\label{binary-outcome}
Y_t \; \sim \; \text{Bernoulli}\left\{p=expit(\gamma_0 + \gamma_1 \cdot t \cdot A_t + \gamma_2 L_t + \gamma_3 Y_{t-1})\right\},
\end{align}
with $A_t$ and $L_t$ generated as in the other settings. For these experiments, the pooled logistic regression model does not serve as a meaningful comparator. This is because its maximum likelihood estimator targets a conditional odds ratio, which is not directly comparable to the risk differences estimated by our approaches. Results are reported in Table \ref{table3}, with bias and coverage evaluated relative to each estimator’s population limit, approximated by using 10000 datasets of size 1 million. As in the previous experiments, the IPW approach produces less precise estimates than G-computation.
\begin{table}[t]
\centering
\begin{threeparttable}
    \caption{Data-generating mechanism for Setting 3.}
    \label{table:dgm_setting3}
    \renewcommand{\arraystretch}{1.2}
    \setlength{\tabcolsep}{8pt}
    \begin{tabular}{lp{10cm}}
        \rowcolor{darkgray}
        \textbf{Component} & \textbf{Description} \\
        Calendar-time data &
        \begin{minipage}{9cm}
        \vspace{4pt}
        \begin{equation*}
        \begin{aligned}
        L_t \; &\sim \; N(\alpha_0 + \alpha_1 L_{t-1} + 
            \alpha_2 A_{t-1},\, 0.2)\\
        A_t \; &\sim \; A_{t-1} + (1-A_{t-1}) \\
        &\hspace{1cm}\cdot \mathrm{Bernoulli}(\mathrm{expit}(
            \beta_0 + \beta_1 L_{t} +  
            \beta_2 Y_{t-1}))\\
        Y_t \; &\sim \; \mathrm{Bernoulli}\left\{p = \mathrm{expit}(
            \gamma_0 + \gamma_1 \cdot t \cdot A_t + 
            \gamma_2 L_t + \gamma_3 Y_{t-1})\right\}
        \end{aligned}
        \end{equation*}
        \vspace{4pt}
        \end{minipage}\\
        \rowcolor{lightgray}
        Visit-time data &
        \begin{minipage}{9cm}
        \vspace{4pt}
        \begin{equation*}
        \begin{aligned}
        L_t \; &\sim \; N(\alpha_0 + \alpha_1 L_{t-1} + 
            \alpha_2 A_{t-1},\, 0.2)\\
        A_t \; &\sim \; A_{t-1} + (1-A_{t-1}) \\
        &\hspace{1cm}\cdot \mathrm{Bernoulli}\{p = \Phi(
            \beta_0 + \beta_1 L_{t} +             \beta_2 Y_{t-1})\}\\
        Y_t \; &\sim \; \mathrm{Bernoulli}\left\{p = \Phi(
            \gamma_0 + \gamma_1 \cdot t \cdot A_t + 
            \gamma_2 L_t + u)\right\}
        \end{aligned}
        \end{equation*}
        \vspace{4pt}
        \end{minipage}
        \newline
        $\Phi$ is the CDF of the standard normal distribution and $u$ is a standard normal noise. This mechanism was chosen to allow analytical computation of the true value of the estimands.
        \\
        Parameters & 
        All parameters as in Setting 2. \\
    \end{tabular}
\end{threeparttable}
\end{table}
\\\\
\noindent
For visit-time data, the binary treatment and outcome were generated according to:
\begin{align}
&A_t \; \sim \;  A_{t-1} + (1-A_{t-1}) \cdot  \text{Bernoulli}\{ p = \Phi(\beta_0 + \beta_1 L_{t} + \beta_2 Y_{t-1})\}\nonumber\\
&Y_t \; \sim \; \text{Bernoulli}\left\{p=\Phi(\gamma_0 + \gamma_1 \cdot t \cdot A_t + \gamma_2 L_t + u)\right\}\label{binary-outcome2},
\end{align}
where $\Phi$ is the cumulative distribution function of the standard normal distribution and $u$ is a standard normal noise. Unlike the previous mechanism \eqref{binary-outcome}, here we removed the direct connection between consecutive outcomes $Y_{t-1} \to Y_t$. Instead, it was replaced with an unmeasured factor $u$ that affects each outcome individually, inducing correlation across time, which is expected in real data. This data-generating process allows us to analytically compute the true value of the baseline-adjusted effect. In contrast, for outcomes generated as in \eqref{binary-outcome}, the population limit of the estimators of $\psi_b$ can only be approximated in large samples, but its interpretation would remain ambiguous due to the possible misspecification of nuisance models.
\begin{table}[b]
\centering
\begin{threeparttable}
    \caption[Simulation results for calendar-time data with binary outcome]{Results for calendar-time data with time-dependent treatment effect and binary outcome: Average estimate, Empirical Bias, Mean Standard Error, Empirical Standard Deviation, 95\% CI Coverage of the proposed estimators and Monte Carlo Errors in parentheses ($\times100$).}
    \label{table3}
    \renewcommand{\arraystretch}{1.2} 
    \setlength{\tabcolsep}{8pt} 
    \begin{tabular}{llccccc}
        \rowcolor{darkgray}
        Data design & Estimator & Estimate & Bias & SE & SD & 95\% Coverage \\
        Calendar-time & Unif. effect IPW & 0.135 & 0.008 (0.10) & 0.030 (0.04) & 0.033 (0.07) & 0.949 (0.70) \\
        & Unif. effect G-comp & 0.135 & 0.008 (0.07) & 0.021 (0.01) & 0.021 (0.05) & 0.921 (0.85) \\
        \rowcolor{lightgray}
        & Elig. effect IPW & 0.133 & -0.002 (0.09) & 0.025 (0.03) & 0.027 (0.06) & 0.940 (0.75) \\
        \rowcolor{lightgray}
        & Elig. effect G-comp & 0.137 & 0.001 (0.06) & 0.022 (0.01) & 0.019 (0.04) & 0.964 (0.59) \\
    \end{tabular}
        \end{threeparttable}
\end{table}
\begin{table}[t]
\centering
\begin{threeparttable}
    \caption[Simulation results for visit-time data with binary outcome]{Results for visit-time data with time-dependent treatment effect and binary outcome: Average estimate, Empirical Bias, Mean Standard Error, Empirical Standard Deviation, 95\% CI Coverage of the proposed estimators and Monte Carlo Errors in parentheses ($\times100$).}
    \label{table4}
    \renewcommand{\arraystretch}{1.2} 
    \setlength{\tabcolsep}{8pt} 
    \begin{tabular}{lllccccc}
        \rowcolor{darkgray}
        Data design & Estimator & Link & Estimate & Bias & SE & SD & 95\% Coverage \\
        Visit-time & B.-adj. IPW & Logit & 0.355 & 0.006 (0.08) & 0.027 (0.04) & 0.026 (0.06) & 0.937 (0.77) \\
        & B.-adj. G-comp & & 0.354 & 0.005 (0.08) & 0.026 (0.02) & 0.025 (0.06) & 0.954 (0.66) \\
        \rowcolor{lightgray}
        & B.-adj. IPW & Probit & 0.353 & 0.004 (0.09) & 0.027 (0.02) & 0.028 (0.06) & 0.941 (0.75) \\
        \rowcolor{lightgray}
        & B.-adj. G-comp & & 0.352 & 0.003 (0.08) & 0.028 (0.01) & 0.026 (0.06) & 0.941 (0.75) \\
    \end{tabular}
\end{threeparttable}
\end{table}
\\\\
\noindent
Table \ref{table4} shows the results obtained by estimators using correctly specified nuisance models (Link: Probit), and by estimators running logistic models for the nuisance parameters (Link: Logit). Logistic models are commonly used in standard practice for modelling the propensity score, but here they are misspecified considering the data-generating process \eqref{binary-outcome2}. The empirical bias and coverage are reported with respect to the analytically computed true value of each estimand. All proposed estimators managed to capture the truth, and achieved low bias and near-nominal coverage. As in the previous settings, G-computation was more efficient. Estimators running nuisance models with logit link produced slightly higher bias, although the difference is negligible. 

\subsection{Working Models for the IPW and G-computation Estimators}
\label{working-models}
Here we present the working models used to construct the estimating equations for the proposed IPW and G-computation estimators. Consider that data were generated for $\tau = 2$ time points.\\\\
\textbf{IPW estimator}\\
For the IPW estimator, the propensity scores are modeled via logistic regressions:
\begin{align*}
&P(A_{t,i}=1|\overline{W}_{t,i},I_{t,i}=1) =  expit(\beta_0 + \beta_1 L_{t} + \beta_2 Y_{t-1}).
\end{align*}
For the setting with binary outcome, the propensity scores were also modeled with a probit link function:
\begin{align*}
&P(A_{t,i}=1|\overline{W}_{t,i},I_{t,i}=1) =  \Phi(\beta_0' + \beta_1' L_{t} + \beta_2' Y_{t-1}).
\end{align*}
For the baseline-adjusted effect, note that the issue of modelling the treatment variable twice using different conditioning sets when modelling $P(I_{t,i}=1|L_{1,i})$ and $P(A_{t,i}=1|\overline{W}_{t,i},I_{t,i}=1)$ (see Appendix S3.5.1 for more details) does not arise, given that we generate data for 2 time points with the data-generating mechanism described above. In fact, at $t=2$ we have to model $P(I_{2,i}=1|L_{1,i})$ which equals $ 1-P(A_{1,i}=1|L_{1,i})$.\\\\
\textbf{G-computation estimator}\\
For the G-computation estimator, conditional outcome means are modeled via linear regressions:
\begin{align*}
&\mathbb{E}(Y_t|A_t = 1,\overline{W}_{t},I_t=1) = \gamma_0 + \gamma_1 L_t + \gamma_2 Y_{t-1}.
\end{align*}
For the baseline-adjusted effect, the predicted outcome means are then regressed on the baseline covariates:
\begin{align*}
&\hat{\mathbb{E}}\left\{\hat{\mathbb{E}}(Y_{t}|A_t=1,I_t=1,\overline{W}_t)|L_1,I_t=1\right\}= \gamma_0' + \gamma_1' L_1.
\end{align*}
The marginal probabilities of eligibility have been computed via their sample mean.

\section{Effect of antimicrobial de-escalation in intensive care}
\label{sec:distributions}
In the data analysis, de-escalation was defined as a change in the antimicrobial prescription downhill a predefined spectrum gradient (from meropenem to piperacillin–tazobactam or amoxicillin–clavulanic acid, and from piperacillin–tazobactam to amoxicillin–clavulanic acid) within 48 hours from the availability of the susceptibility results. The study population was restricted to patients older than 18 years, who received empirical meropenem or piperacillin–tazobactam for at least 48 hours during ICU stays of $\geq 96$ hours, had confirmed bacterial etiology with susceptibility data, and were eligible for de-escalation under hospital protocols. The final dataset included data for 241 unique patients, extracted from the COSARA electronic platform \supercite{steurbaut2012cosara} as part of routine clinical care. The data were organized with a visit-time design, with 241 eligible patients in visit 1 and 160 in visit 2. Time 0 (time of inclusion) for visit 1 was defined as the moment when the in vitro susceptibility of the causative pathogen became available. Patients assigned to the control strategy in visit 1 and still hospitalized at the ICU and on antimicrobial (AB) therapy 24 hours after inclusion in the first trial were considered eligible for inclusion in visit 2. Time 0 for visit 2 was hence defined as 24 hours after inclusion in the first trial. 
\begin{table}[H]
\centering
\begin{threeparttable}
\caption{Summary statistics of the outcomes at visit 1 by de-escalation group.}
\label{tab:visit1_outcomes}
\begin{tabular}{lccc}
\rowcolor{darkgray}
Outcome & Joint (n=241) & Continuation (n=169) & De-escalation (n=72)\\
\hline
\rowcolor{lightgray} Hospital survival (\%) & 73.03 & 72.19 & 75.00 \\
Hospital time (days) & 41.20 (63.49) & 44.28 (71.32) & 33.97 (38.90) \\
\rowcolor{lightgray} ICU time (days) & 10.10 (15.17) & 10.54 (15.63) & 9.06 (14.08) \\
Total AM (days) & 9.27 (19.26) & 9.87 (21.53) & 7.88 (12.39) \\
\hline
\end{tabular}
\begin{tablenotes}
       \item  Continuous variables are presented as mean (SD); categorical variables as \%. 
       \item Abbreviations: Hospital Time denotes duration of hospital stay; ICU Time denotes duration of ICU stay; Total AM denotes total antimicrobial consumption.
\end{tablenotes}
\end{threeparttable}
\end{table}
\begin{table}[H]
\centering
\begin{threeparttable}
\caption{Summary statistics of the outcomes at visit 2 by de-escalation group.}
\label{tab:visit2_outcomes}
\begin{tabular}{lccc}
\rowcolor{darkgray}
Outcome & Joint (n=160)& Continuation (n=127)& De-escalation (n=33)\\
\hline
\rowcolor{lightgray} Hospital survival (\%) & 71.88 & 68.50 & 84.85 \\
Hospital time (days) & 43.27 (72.38) & 44.52 (79.35) & 38.47 (34.80) \\
\rowcolor{lightgray} ICU time (days) & 10.12 (15.87) & 10.15 (16.83) & 10.00 (11.65) \\
Total AM (days) & 9.35 (21.95) & 9.32 (23.52) & 9.45 (14.73) \\
\hline
\end{tabular}
\end{threeparttable}
\end{table}
The dataset included 14 covariates that were chosen for their clinical relevance to the causal question: demographics (age, sex), urgent admission, admission to medical/surgical ward, severity scores (SAPS II, APACHE II, daily SOFA II), presence of a respiratory infection, probability of infection, presence of bacteraemia, presence of sepsis, days in hospital before inclusion, laboratory shift during which the microbiology samples were processed, rank of the antimicrobials prescribed during the empirical phase. Daily SOFA II scores were recorded throughout the ICU stay. Since the timing of susceptibility reporting is not stored in COSARA, availability times were imputed from sample timestamps using standard laboratory turnaround intervals 
(48 or 72 hours). Antimicrobial dosing regimens were obtained from previously compiled records. All other variables were collected at the time of admission. Tables~\ref{tab:visit1_outcomes}--\ref{tab:visit2_summary} illustrate how the distributions of outcomes and covariates evolve across the two time points. The largest variation was registered for the presence of bacteraemia and for admission to the medical/surgical ward. Future work will include an additional analysis of the same dataset and further information regarding data collection procedures and preprocessing.
\begin{table}[H]
\centering
\begin{threeparttable}
\caption{Summary statistics of the covariates at visit 1 by de-escalation group.}
\label{tab:visit1_summary}
\begin{tabular}{lccc}
\rowcolor{darkgray}
Covariate & Joint (n=241) & Continuation (n=169) & De-escalation (n=72)\\
\hline
\rowcolor{lightgray} AdmType (\%) & 83.40 & 83.43 & 83.33 \\
Age (years) & 60.45 (15.71) & 60.33 (16.61) & 60.75 (13.45) \\
\rowcolor{lightgray} APACHE II score & 23.97 (8.01) & 23.67 (8.41) & 24.67 (6.97) \\
Bacteraemia (\%) & 23.24 & 17.75 & 36.11 \\
\rowcolor{lightgray} $\Delta S$ (corrected) & 0.41 (2.01) & 0.28 (1.88) & 0.72 (2.28) \\
Empirical rank (\%) & 13.28 & 13.02 & 13.89 \\
\rowcolor{lightgray} Infection probability (\%) & 0.78 & 0.74 & 0.88  \\
Lab shift (\%) & 20.33 & 23.08 & 13.89 \\
\rowcolor{lightgray} Respiratory (\%) & 49.79 & 50.89 & 47.22 \\
SAPS II score & 59.53 (19.79) & 59.22 (21.05) & 60.25 (16.56) \\
\rowcolor{lightgray} Sepsis (\%) & 48.96 & 44.97 & 58.33 \\
Sex\tnote{a} (\%) & 72.61 & 74.56 & 68.06 \\
\rowcolor{lightgray} Time before (days) & 8.66 (8.75) & 9.70 (9.49) & 6.21 (6.09) \\
Ward (\%) & 58.51 & 62.13 & 50.00 \\
\hline
\end{tabular}
\begin{tablenotes}
       \item [a] Sex is defined as 1 for males, 0 for females.
       \item Abbreviations: AdmType denotes urgent admission or not; $\Delta S$ (corrected) denotes the corrected daily change in SOFA II score; Empirical rank denotes the rank of the empirical antimicrobials (1 for rank 3, i.e. meropenem, 0 for rank 2 i.e. Piperacillin + Tazobactam); Lab shift denotes the laboratory shift of the microbiology samples (1 for 18:30 - 00:00, 0 for 00:00 - 18:30); Respiratory denotes the presence of a respiratory infection; Time before denotes the amount of days in hospital before inclusion; Ward denotes admission to medical/surgical ward.
\end{tablenotes}
\end{threeparttable}
\end{table}
\begin{table}[H]
\centering
\begin{threeparttable}
\caption{Summary statistics of the covariates at visit 2 by de-escalation group.}
\label{tab:visit2_summary}
\begin{tabular}{lccc}
\rowcolor{darkgray}
Covariate & Joint (n=160)& Continuation (n=127)& De-escalation (n=33)\\
\hline
\rowcolor{lightgray} AdmType (\%) & 84.38 & 86.61 & 75.76 \\
Age (years) & 59.94 (16.81) & 58.55 (17.52) & 65.27 (12.62) \\
\rowcolor{lightgray} APACHE II score & 23.86 (8.33) & 23.71 (7.83) & 24.42 (10.17) \\
Bacteraemia (\%) & 17.50 & 17.32 & 18.18 \\
\rowcolor{lightgray} $\Delta S$ (corrected) & 0.60 (1.73) & 0.55 (1.77) & 0.79 (1.60) \\
Empirical rank (\%) & 13.75 & 14.96 & 9.09 \\
\rowcolor{lightgray} Infection probability (\%) & 0.74  & 0.75  & 0.70  \\
Lab shift (\%) & 22.50 & 25.98 & 9.09 \\
\rowcolor{lightgray} Respiratory (\%) & 51.25 & 48.82 & 60.61 \\
SAPS II score & 59.74 (20.80) & 59.91 (20.12) & 59.09 (23.54) \\
\rowcolor{lightgray} Sepsis (\%) & 46.25 & 47.24 & 42.42 \\
Sex (\%) & 74.38 & 75.59 & 69.70 \\
\rowcolor{lightgray} Time before (days) & 10.55 (9.59) & 10.86 (9.74) & 9.34 (9.03) \\
Ward (\%) & 63.13 & 66.14 & 51.52 \\
\hline
\end{tabular}
\end{threeparttable}
\end{table}

\subsection{Validity of causal assumptions in the data analysis}
\label{sec:validity}
\begin{figure}[H]
\centering
\begin{subfigure}{0.37\textwidth}
    \includegraphics[width=\textwidth]{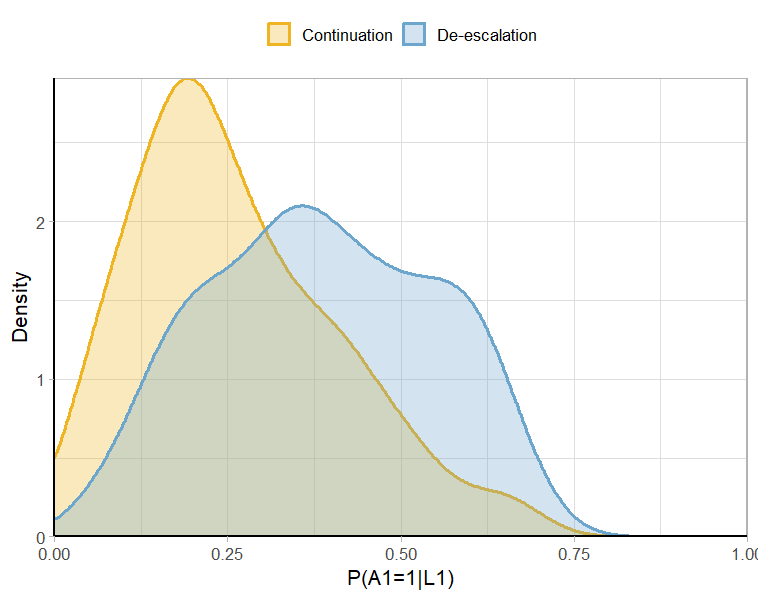}
    \caption{}
    \label{fig:ps1}
\end{subfigure}
\hspace{1cm}
\begin{subfigure}{0.37\textwidth}
    \includegraphics[width=\textwidth]{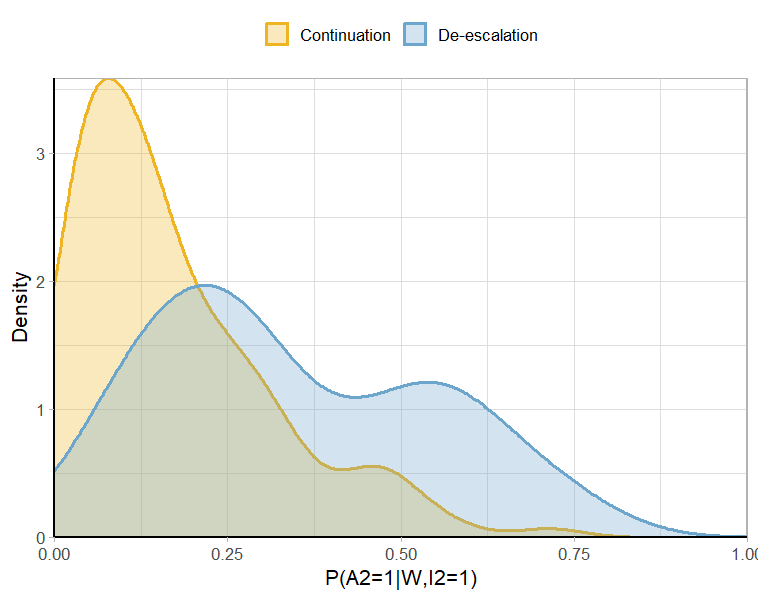}
    \caption{}
    \label{fig:ps3-hosp_survival}
\end{subfigure}
\caption[Distributions of the estimated propensity scores at visits 1 and 2]{The plots show the distributions of the estimated probability of receiving treatment truncated at the 95th percentile at visit 1 (a), and at visit 2 (b). The yellow curve is for $A_t=0$, the blue curve for $A_t=1$.}
\label{fig:ps}\vspace{0.5cm}
\RaggedRight\textit{Alt text:} Two density plots of estimated propensity scores, truncated at the 95th percentile, comparing patients who continued treatment (yellow) versus those de-escalated (blue). Panel (a), visit 1: yellow peaks sharply at a low probability (around 0.15) and declines quickly; blue is broader and shifted right, peaking around 0.3, with substantial overlap between groups. Panel (b), visit 2: yellow peaks even more sharply near 0.05; blue is bimodal, with a first peak around 0.15 and a smaller second peak around 0.5, extending further right than yellow, with overlap concentrated at lower probability values.
\end{figure}

The application of the proposed estimators in the data analysis relies on the standard causal assumptions of consistency, positivity and sequential exchangeability. Sequential exchangeability requires that all common causes of de-escalation and each of the outcomes were measured. In this dataset, key confounders such as severity of illness, infection site and antimicrobial history were included, but some unmeasured factors, such as the clinician's judgment of patients' conditions, could still influence both treatment and outcome. Because de-escalation is typically initiated when a patient is judged to be improving, the potential bias would therefore make de-escalation appear more beneficial than it actually is. However, we expect the bias to be modest, since the measured confounders, such as severity, infection type, and antimicrobial history, are likely to capture a substantial part of the clinician's assessment.
\\\\
Figure \ref{fig:ps} gives insights on the assumption on positivity of treatment assignment, showing the distributions of the estimated propensity scores for both time points, truncated at the 95th percentile. The primary and secondary outcomes are in the conditioning set of the propensity score at $t=2$. The plots display reasonable overlap between treatment groups, though some values close to zero were observed, particularly for the second time point, which can be a reason for the larger standard errors of the IPW estimators. However, the corresponding weights were not extreme, the largest one being 47.9.
\noindent
Regarding eligibility, consider that all patients have probability 1 of being eligible at visit 1, while $I_2=1-A_1$ at visit 2. The distribution of the estimated probability of being eligible at visit 2 can thus be read from Figure S3(a), which doesn't imply the presence of too extreme values.

\subsection{Target populations of the proposed estimands}
The proposed estimands in Section \textit{Model-free estimands for target trial emulation} of the main text imply different target populations, which can be characterised by the table of the covariate distributions aggregated according to the estimand's weighting scheme. In practice, (weighted) descriptive statistics for the implied target population can be computed on the pooled dataset of all eligible patient-time units, where each unit is weighted according to the estimand's scheme. 
\\\\
The proposed estimands differ in how the weighting in the pooled dataset is performed. For $\psi_e$ each patient-time unit receives $w_e=1$, so they contribute equally;  for $\psi_u$, each unit receives weight $w_u=1/n_t$, where $n_t$ is the number of patient-time units in trial $t$, so that all trials contribute equally regardless of their size. For $\psi_b$, the target population is the one defined at baseline.
\\\\
Table S9 describes the target populations implied by each estimand for the dataset used in the data analysis. Consider that the only time-varying variables in the dataset are $\Delta S$ (corrected) and Time before. All other covariates are recorded at baseline only and remain constant across visits, but since eligibility changes across visits, they also contribute to defining different target populations. Continuous variables are presented as mean (SD); categorical variables as \%.
\begin{table}[t]
\centering
\caption[Target populations implied by the proposed estimands]{Target populations implied by the proposed estimands 
for the time-varying covariates.}
\begin{minipage}{0.3\textwidth}
\centering
\begin{threeparttable}
\caption*{$\psi_u$}
\begin{tabular}{lc}
\rowcolor{darkgray}
Covariate & Value \\
\hline
\rowcolor{lightgray} AdmType (\%) & 83.89 \\
Age (years) & 60.20 (16.26) \\
\rowcolor{lightgray} APACHE II score & 23.92 (8.17) \\
Bacteraemia (\%) & 20.37 \\
\rowcolor{lightgray} $\Delta S$ (corrected) & 0.50 (1.88) \\
Empirical rank (\%) & 13.52 \\
\rowcolor{lightgray} Infection prob. (\%) & 0.76 \\
Lab shift (\%) & 21.42 \\
\rowcolor{lightgray} Respiratory (\%) & 50.52 \\
SAPS II score & 59.63 (20.30) \\
\rowcolor{lightgray} Sepsis (\%) & 47.60 \\
Sex (\%) & 73.50 \\
\rowcolor{lightgray} Time before (days) & 9.60 (9.20) \\
Ward (\%) & 60.82 \\
\hline
\end{tabular}
\end{threeparttable}
\end{minipage}
\hfill
\begin{minipage}{0.3\textwidth}
\centering
\begin{threeparttable}
\caption*{$\psi_e$}
\begin{tabular}{lc}
\rowcolor{darkgray}
Covariate & Value \\
\hline
\rowcolor{lightgray} AdmType (\%) & 83.79 \\
Age (years) & 60.25 (16.14) \\
\rowcolor{lightgray} APACHE II score & 23.93 (8.13) \\
Bacteraemia (\%) & 20.95 \\
\rowcolor{lightgray} $\Delta S$ (corrected) & 0.49 (1.90) \\
Empirical rank (\%) & 13.47 \\
\rowcolor{lightgray} Infection prob. (\%) & 0.76 \\
Lab shift (\%) & 21.20 \\
\rowcolor{lightgray} Respiratory (\%) & 50.37 \\
SAPS II score & 59.61 (20.20) \\
\rowcolor{lightgray} Sepsis (\%) & 47.88 \\
Sex (\%) & 73.32 \\
\rowcolor{lightgray} Time before (days) & 9.41 (9.12) \\
Ward (\%) & 60.35 \\
\hline
\end{tabular}
\end{threeparttable}
\end{minipage}
\hfill
\begin{minipage}{0.3\textwidth}
\centering
\begin{threeparttable}
\caption*{$\psi_b$}
\begin{tabular}{lc}
\rowcolor{darkgray}
Covariate & Value \\
\hline
\rowcolor{lightgray} AdmType (\%) & 83.40 \\
Age (years) & 60.45 (15.71) \\
\rowcolor{lightgray} APACHE II score & 23.97 (8.01)\\
Bacteraemia (\%) & 23.24 \\
\rowcolor{lightgray} $\Delta S$ (corrected) & 0.41 (2.01) \\
Empirical rank (\%) & 13.28  \\
\rowcolor{lightgray} Infection prob. (\%) & 0.78  \\
Lab shift (\%) & 20.33 \\
\rowcolor{lightgray} Respiratory (\%) & 49.79  \\
SAPS II score & 59.53 (19.79) \\
\rowcolor{lightgray} Sepsis (\%) & 48.96 \\
Sex (\%) & 72.61 \\
\rowcolor{lightgray} Time before (days) & 8.66 (8.75)  \\
Ward (\%) & 58.51  \\
\hline
\end{tabular}
\end{threeparttable}
\end{minipage}
\end{table}

\subsection{Results on the Log-Odds Scale for Hospital Survival}
\label{section:mle}
We evaluated the proposed estimators on the log-odds scale for the binary hospital survival outcome of the data analysis.
The corresponding estimands are the logit-transformed analogs of those defined on the risk difference scale in Section \textit{Model-free estimands for target trial emulation}:
\begin{enumerate}
    \item \textbf{Uniformly weighted effect in a random trial}\\
    \begin{align*}
        \psi_u 
        = & \; g\left\{\frac{1}{\tau}\sum_{t=1}^{\tau} \mathbb{E}(Y^1_t|I_t=1)\right\} - g\left\{\frac{1}{\tau}\sum_{t=1}^{\tau} \mathbb{E}(Y^0_t|I_t=1)\right\}
    \end{align*}
    \item \textbf{Eligibility-weighted effect in a random trial}\\
    \begin{align*}
        \psi_e = g\left\{\sum_{t=1}^{\tau}\mathbb{E}(Y^1_t|I_t=1)\frac{P(I_t=1)}{\sum_{j=1}^{\tau} P(I_j=1)}\right\} - g\left\{\sum_{t=1}^{\tau}\mathbb{E}(Y^0_t|I_t=1)\frac{P(I_t=1)}{\sum_{j=1}^{\tau} P(I_j=1)}\right\}
    \end{align*}\\
    \item \textbf{Baseline-adjusted effect}\\
    \begin{flalign*}
        \psi_b = g\left[\frac{1}{\tau}\sum^\tau_{t=1}\mathbb{E}\left\{\mathbb{E}\left(Y^1_{t}|I_t=1,L_1\right)\right\}\right]-g\left[\frac{1}{\tau}\sum^\tau_{t=1}\mathbb{E}\left\{\mathbb{E}\left(Y^0_{t}|I_t=1,L_1\right)\right\}\right]
    \end{flalign*}
\end{enumerate}
where the link function $g(\cdot)$ is the logit function $logit(p)=ln\left(\frac{p}{1-p}\right)$. 
While these effect measures carry interpretability issues, as we explain in Section \textbf{Noncollapsible effect measures} of the main text, it is possible to compare their corresponding estimators to standard MLEs.
Identification and estimation follow the same logic as for the original estimands, with the addition of the logit link. IPW and G-computation estimators were implemented, truncating IPW weights at the 95th percentile to balance bias and variance. Standard errors were computed using the \textit{geex} package to account for clones. As shown in Table~\ref{table6}, all estimators produce odds ratios close to 1, with confidence intervals including unity, suggesting little evidence for the effect of antimicrobial de-escalation on hospital survival, consistent with the results on the risk difference scale in Section \textit{Effect of antimicrobial de-escalation in intensive care} of the main text. As expected, IPW estimates had larger standard errors, reflecting higher variability.
\begin{table}[b]
\centering
\begin{threeparttable}
    \caption[Odds ratios estimated by the proposed estimators]{Odds ratios (log odds in brackets) obtained by the proposed estimators. Estimated Odds Ratio, Standard Error, lower and upper bounds of the 95\% Confidence Intervals.}
    \label{table6}
    \renewcommand{\arraystretch}{1.2} 
    \setlength{\tabcolsep}{8pt} 
    \begin{tabular}{lcccc}
        \rowcolor{darkgray}
        Estimator & Odds Ratio (log Odds) & SE & CI lower & CI upper \\
        Unif. effect IPW &  1.19 (0.17)  &  0.63 &  0.34 ($-1.07$)  & 4.12  (1.42)\\
        Unif. effect G-comp& 1.01 (0.01) & 0.11 & 0.81 ($-0.21$) &  1.27 (0.24) \\
        \rowcolor{lightgray}
        Elig. effect IPW&  1.22 (0.20)  & 0.59  &   0.38 ($-0.97$) & 3.95 (1.37)\\
        \rowcolor{lightgray}
        Elig. effect G-comp&   1.01 (0.01)  & 0.20 &  0.68 ( $-0.38$) & 1.51 (0.41)\\
        Baseline-adj. IPW&  1.12 (0.11) & 0.55  &  0.38 ($-0.97$) & 3.34 (1.20)\\
        Baseline-adj. G-comp&  1.01 (0.01) & 0.12  & 0.80 ($-0.22$) & 1.26 (0.23)\\
    \end{tabular}
\end{threeparttable}
\end{table}
\begin{table}[t]
\centering
\begin{threeparttable}
    \caption[Odds ratios estimated by pooled and time-specific logistic regressions]{Odds ratios (log odds in brackets) obtained by pooled and time-specific logistic regressions. Estimated Odds Ratios, Standard Error, lower and upper bounds of the 95\% Confidence Intervals.}
    \label{table7}
    \renewcommand{\arraystretch}{1.2} 
    \setlength{\tabcolsep}{8pt} 
    \begin{tabular}{lcccc}
        \rowcolor{darkgray}
        Estimator & Odds Ratio (log Odds) & SE & CI lower & CI upper \\
        Pooled MLE & 2.38 (0.87) &  0.32  &  1.26 (0.23) & 4.48 (1.50)\\
        Stepwise Pooled MLE & 2.53 (0.93) &  0.33  & 1.33 (0.29) & 4.82 (1.57)\\
        \rowcolor{lightgray}
        MLE $t = 1$ & 1.75 (0.56) &  0.37  & 0.84 ($-0.17$) & 3.65 (1.29)\\
        \rowcolor{lightgray}
        Stepwise MLE $t = 1$ & 1.68 (0.52) &  0.36  & 0.83 ($-0.19$) & 3.41 (1.23)\\
        \rowcolor{lightgray}
        MLE $t = 2$ & 5.72 (1.74) &  0.68  & 1.52 (0.42) & 21.59 (3.07)\\
        \rowcolor{lightgray}
        Stepwise MLE $t = 2$ & 5.7 (1.74) &  0.62  & 1.68 (0.52) & 19.37 (2.96)\\    
    \end{tabular}
    \end{threeparttable}
\end{table}
\\\\
\noindent
For comparison, pooled and time-specific logistic regressions were also fitted (Table~\ref{table7}). The pooled model, which included visit time as a covariate, produced a large estimated odds ratio of 2.38. Time-specific regressions indicated apparent effect heterogeneity: at $t=1$, the estimated odds ratio was about 1.7 (not significant), while at $t=2$ it increased to roughly 5.7, with wide confidence intervals excluding the null. The large difference between time-specific effects raises interpretational challenges. Similarly to the results in Section \textit{Noncollapsible effect measures} of the main text, such variation may be driven by noncollapsibility, rather than true effect modification, because the distributions of the covariates change across time points (see Tables~\ref{tab:visit1_summary}--\ref{tab:visit2_summary}). The inflated estimates of the MLEs at both visits compared to the proposed estimators would thus be a consequence of noncollapsibility. Moreover, the pooled odds ratio represents a weighted mixture of time-specific effects across heterogeneous populations, with unclear causal interpretation. As an additional investigation, we ran G-computation on the pooled data in long format, producing an estimated odds ratio of 1.74 (95\% CI: 1.50 to 2.01). This suggests that the large odds ratio of the pooled MLE may be partly due to potential outcome model misspecification when pooling the data.
\\\\
Due to the large number of covariates (14) compared to the number of patients (241 at $t=1$ and 160 at $t=2$), we refined the logistic model via stepwise AIC selection, but this did not affect the conclusions. Finally, note that such logistic analyses yield effect estimates that are very difficult to interpret: the odds of surviving if patients are de-escalated, conditional on the covariates. This is not entirely comparable to the marginal estimates of the proposed estimators.

\printbibliography
\end{refsection}

\end{document}